\documentclass[prx, aps, twocolumn, superscriptaddress, reprint, nobalancelastpage, nofootinbib]{revtex4-2}

\usepackage[pdftex]{graphicx}
\usepackage{hyperref}
\usepackage{amsmath}
\usepackage{physics}
\usepackage{xcolor}
\usepackage{nicefrac}
\usepackage{siunitx}
\usepackage{bbold}

\newcommand{\MS}{{M{\o}lmer-S{\o}rensen }}

\begin{document}

\title{Demonstration of fault-tolerant universal quantum gate operations}

\author{Lukas Postler}
\affiliation{Institut f\"{u}r Experimentalphysik, Universit\"{a}t Innsbruck, Innsbruck, Austria}

\author{Sascha Heu\ss en}
\affiliation{Institute for Quantum Information, RWTH Aachen University, Aachen, Germany}
\affiliation{Institute for Theoretical Nanoelectronics (PGI-2), Forschungszentrum J\"{u}lich, J\"{u}lich, Germany}

\author{Ivan Pogorelov}
\affiliation{Institut f\"{u}r Experimentalphysik, Universit\"{a}t Innsbruck, Innsbruck, Austria}

\author{Manuel Rispler}
\affiliation{Institute for Quantum Information, RWTH Aachen University, Aachen, Germany}
\affiliation{Institute for Theoretical Nanoelectronics (PGI-2), Forschungszentrum J\"{u}lich, J\"{u}lich, Germany}

\author{Thomas Feldker}
\affiliation{Institut f\"{u}r Experimentalphysik, Universit\"{a}t Innsbruck, Innsbruck, Austria}
\affiliation{Alpine Quantum Technologies GmbH, Innsbruck, Austria}

\author{Michael Meth}
\affiliation{Institut f\"{u}r Experimentalphysik, Universit\"{a}t Innsbruck, Innsbruck, Austria}

\author{Christian D. Marciniak}
\affiliation{Institut f\"{u}r Experimentalphysik, Universit\"{a}t Innsbruck, Innsbruck, Austria}

\author{Roman Stricker}
\affiliation{Institut f\"{u}r Experimentalphysik, Universit\"{a}t Innsbruck, Innsbruck, Austria}

\author{Martin Ringbauer}
\affiliation{Institut f\"{u}r Experimentalphysik, Universit\"{a}t Innsbruck, Innsbruck, Austria}

\author{Rainer Blatt}
\affiliation{Institut f\"{u}r Experimentalphysik, Universit\"{a}t Innsbruck, Innsbruck, Austria}
\affiliation{Institut f\"{u}r Quantenoptik und Quanteninformation, \"{O}sterreichische Akademie der Wissenschaften, Innsbruck, Austria}

\author{Philipp Schindler}
\affiliation{Institut f\"{u}r Experimentalphysik, Universit\"{a}t Innsbruck, Innsbruck, Austria}

\author{Markus M\"{u}ller}
\affiliation{Institute for Quantum Information, RWTH Aachen University, Aachen, Germany}
\affiliation{Institute for Theoretical Nanoelectronics (PGI-2), Forschungszentrum J\"{u}lich, J\"{u}lich, Germany}

\author{Thomas Monz}
\affiliation{Institut f\"{u}r Experimentalphysik, Universit\"{a}t Innsbruck, Innsbruck, Austria}
\affiliation{Alpine Quantum Technologies GmbH, Innsbruck, Austria}

\begin{abstract}

Quantum computers can be protected from noise by encoding the logical quantum information redundantly into multiple qubits using error correcting codes. 
When manipulating the logical quantum states, it is imperative that errors caused by imperfect operations do not spread uncontrollably through the quantum register. This requires that all operations on the quantum register obey a fault-tolerant circuit design which, in general, increases the complexity of the implementation.
Here, we demonstrate a fault-tolerant universal set of gates on two logical qubits in a trapped-ion quantum computer. In particular, we make use of the recently introduced paradigm of flag fault tolerance, where the absence or presence of dangerous errors is heralded by usage of few ancillary 'flag' qubits. 
We perform a logical two-qubit \mbox{CNOT-gate} between two instances of the seven qubit color code, and we also fault-tolerantly prepare a logical magic state. We then realize a fault-tolerant logical T-gate by injecting the magic state via teleportation from one logical qubit onto the other. We observe the hallmark feature of fault tolerance, a superior performance compared to a non-fault-tolerant implementation. In combination with recently demonstrated repeated quantum error correction cycles these results open the door to error-corrected universal quantum computation.

\end{abstract}

\maketitle

\section{Introduction}

\begin{figure}
    \includegraphics[width=89mm]{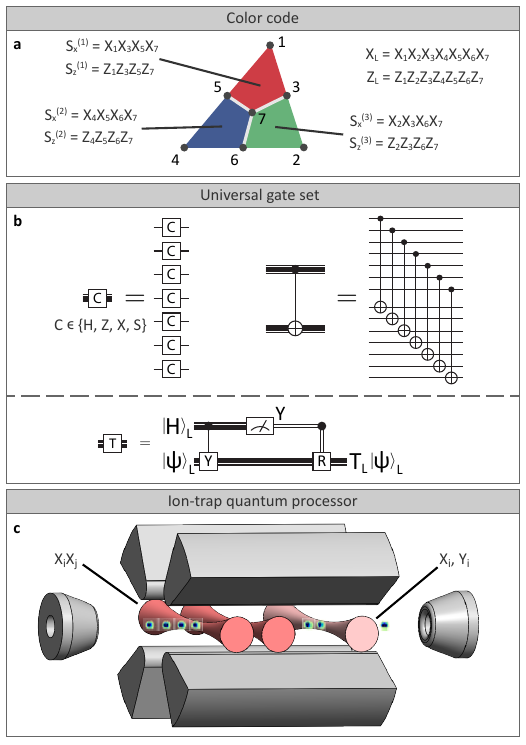}
    \caption{QEC code, logical gates and experimental system. (a) Seven-qubit color code encoding one logical qubit in seven physical qubits. The six weight-4 operators $\{S_x^{(i)},S_z^{(i)}\}$, $i=1,2,3$, are the stabilizer generators and the weight-7 operators are logical operators $Z_L$ and $X_L$.
    (b) Universal gate set consisting of Clifford gates (above dashed line) and the \mbox{T-gate}. Whereas the Clifford group is transversal in the color code, magic state injection can be used to realize the fault-tolerant \mbox{T-gate}. The magic state $\ket{H}_L$ is prepared fault-tolerantly and subsequently teleported onto the target qubit in an arbitrary state $\ket{\Psi}_L$, effectively implementing a \mbox{T-gate} on the target qubit.
    (c) Schematic 3D model of the ion-trap quantum processor. Single and any pair $i,j$ of ions can be addressed simultaneously by steerable tightly focused laser beams. This enables entangling (darker shaded beams) but also single-qubit (lighter) gates.}
    \label{fig:FT_concepts}
\end{figure}

Quantum computers promise to efficiently solve important computational tasks that are beyond the capabilities of classical computers, such as prime factorization or the simulation of complex quantum systems~\cite{Shor1997, Feynman1982}. A digital quantum computer will offer a native gate set, which is comprised of the operations that can be physically executed in hardware. Remarkably, finite sets of native gates are sufficient to compose \emph{any} operation to an arbitrary desired precision, rendering such gate sets \emph{universal}~\cite{Nielsen2010}. A fundamental challenge is to keep the quantum computation coherent, while all components of the quantum computer such as physical qubits, gate operations and measurements are inherently prone to errors. This roadblock towards large-scale quantum computation can be lifted with the tools of \emph{quantum fault tolerance}~\cite{Shor1996, Preskill1998,Aliferis2006,Terhal2015}. The central idea is to use quantum error correction (QEC) codes, where many physical qubits together comprise so-called logical qubits such that the logical information is distributed non-locally and thereby protected from decoherence and errors due to finite control accuracy. By ingenuity of code design, it thus becomes possible to suppress logical decoherence arbitrarily by adding redundancy, once the physical noise level falls below some threshold~\cite{Aharonov2008}. Arbitrary logical quantum computation demands that a universal logical gate set has to be synthesized from physical gates, which presents new challenges: to prevent previously localized errors from spreading over the entire qubit register and destroying the computation, logical gates have to be designed with fault tolerance guarantees: For QEC codes with the potential to correct at least one arbitrary single error, this means that a single error occurring at any location (initialization, gate or measurement) in a particular circuit may under no circumstances turn into a non-correctable error on two or more qubits. When assuming for simplicity that every location has some error probability $p$, the logical failure rate without fault tolerance will scale as $p_L \propto Np$ with $N$ the number of error locations in the circuit that lead to a logical error. While adding further gates and qubits for fault tolerance increases the number of circuit locations, the logical failure rate will now scale with $p_L \propto N' p^2$, i.e.~it is quadratically suppressed in $p$, where $N'$ now denotes the number of pairs of locations where two errors lead to a logical error. This entails one of the hallmark features of fault-tolerant (FT) implementations: despite adding more (noisy) qubit and gates, the quality of the encoded information can be improved, if the physical noise level is sufficiently low. Certain QEC codes facilitate a fault-tolerant implementation of some gates by acting on all physical qubits individually -- called a transversal logical gate. However, a universal gate set with all gates having a transversal unitary implementation is forbidden by a no-go theorem~\cite{Eastin2009}. This leads to the difficulty that to reach universal fault-tolerant computation at least one logical gate must be implemented by other means, such as magic state injection~\cite{Bravyi2005} or code switching~\cite{Paetznick2013}. Fulfilling fault tolerance requirements for these approaches typically implies a substantial resource overhead~\cite{Beverland2021}.

The growing experimental effort towards fault-tolerant quantum computation has seen tremendous advances: Non-fault-tolerant logical state preparation and transversal single-qubit logical gate operations were shown in a seven qubit experiment with trapped ions~\cite{Nigg2014}. State preparation of a topological surface code state~\cite{Satzinger2021}, repetitive stabilizer measurements in an error-detecting surface code~\cite{Andersen2020,Marques2021} and exponential error suppression in a repetition code~\cite{Chen2021} have been demonstrated in superconducting architectures. Shortly following a theory proposal towards demonstrations of fault tolerance in small systems~\cite{Gottesman2016}, experiments showed state preparation using error detection codes and post-selection~\cite{Takita2017, Vuillot2018, Linke2017}, and recently error correction for fault-tolerant state preparation and a fault-tolerant logical single-qubit Clifford gate~\cite{Egan2021}. Theory works have substantially reduced the resource requirements for fault tolerance by the concept of flag fault tolerance~\cite{Chao2018,Chamberland2018,Chamberland2019,Chao2020,Reichardt2020}. Here, dedicated auxiliary qubits are introduced, which signal the presence of dangerous errors. This concept was used to demonstrate fault-tolerant operation of the five-qubit code in an NV center-based quantum processor~\cite{Abobeih2021}, and fault-tolerant parity check measurements~\cite{Hilder2021} and repetitive rounds of fault-tolerant QEC cycles~\cite{Ryan-Anderson2021} with trapped ions. In the present work, alongside the operation of single-qubit logical Clifford gates, we demonstrate the fault-tolerant implementation of a logical \mbox{CNOT-gate} between two logical color code qubits, thereby realizing the entire Clifford group fault-tolerantly. Since this is all one can hope for regarding transversal implementations, to obtain the non-Clifford gate required for universality, we amend the gate set by a \mbox{T-gate}. For this, in a first step we prepare a magic state fault-tolerantly by the use of flag qubits, as proposed in~\cite{Chamberland2019}. Finally, using this fault-tolerantly prepared magic state and the transversal logical \mbox{CNOT-gate}, we perform fault-tolerant magic state injection, thus demonstrating a universal fault-tolerant gate set.

Matching experimental capabilities on the one hand with fault tolerance requirements on the other hand makes the seven-qubit color code an attractive candidate system for the FT operation of a logical qubit. This QEC code, illustrated in Fig.~\ref{fig:FT_concepts}a, is the smallest member of the code family of topological 2D color codes~\cite{Bombin2006} and is also known as the Steane code~\cite{Steane1996}. It hosts one logical qubit and can be formulated as a stabilizer code on seven physical qubits, with logical states encoded in the joint $+1$-eigenspace of six weight-4 Pauli operators. The logical operators are $ X_L=X^{\otimes 7}$ and $Z_L=Z^{\otimes 7}$, which are stabilizer-equivalent to weight three operators, e.g.~$Z_L\simeq Z_1Z_2Z_3$, rendering it a distance 3 code. This entails that all single-qubit errors can be corrected, but weight-2 errors on the code will lead to logical failures. Besides a transversal \mbox{CNOT-gate}, which is common to all QEC codes of the CSS (Calderbank-Shor-Steane) code family, remarkably it also admits the transversal implementation of the Hadamard gate $H$ and the phase gate $S$. Consequently the entire Clifford group can be implemented transversally (see Fig.~\ref{fig:FT_concepts}b)~\cite{Nielsen2010}. The required magic resource state to enable \mbox{T-gate} injection can be prepared fault-tolerantly thanks to a recently proposed protocol~\cite{Chamberland2019}.

The experiments presented in this work have been performed in a 16-qubit ion-trap quantum information processor~\cite{Pogorelov2021} shown schematically in Fig.~\ref{fig:FT_concepts}c. The native gate set supported by this architecture is composed of entangling \MS operations~\cite{Sorensen2000} effectively implementing \mbox{XX-rotations} with an error rate of $p_2 = 2.5 \times 10^{-2}$, single-qubit rotations around an arbitrary axis in the equatorial plane of the Bloch sphere with an error rate of $p_1 = 5 \times 10^{-3}$ and Z-rotations in software. Error rates for state initialization and measurement are estimated at $p_i, p_m = 3 \times 10^{-3}$. A more detailed discussion on experimental methods can be found in Sec.~\ref{sec:methods_exp}. For better readability, all circuits shown in this work are provided in standard \mbox{CNOT-gates}, as these are equivalent to \mbox{XX-gates} up to local Clifford operations~\cite{Nebendahl2009}. Experimental results are accompanied by Monte Carlo simulations, for which we model imperfections as uniform depolarizing noise on single-qubit gates, initialization and measurement as well as two-qubit gates with independent physical error rates $p_1,\,p_i,\,p_m,\,p_2$, respectively, as described in Sec.~\ref{sec:methods_theory}.

\section{Initializing and characterizing the logical qubit}

\begin{figure*}
    \centering
    \includegraphics[width=183mm]{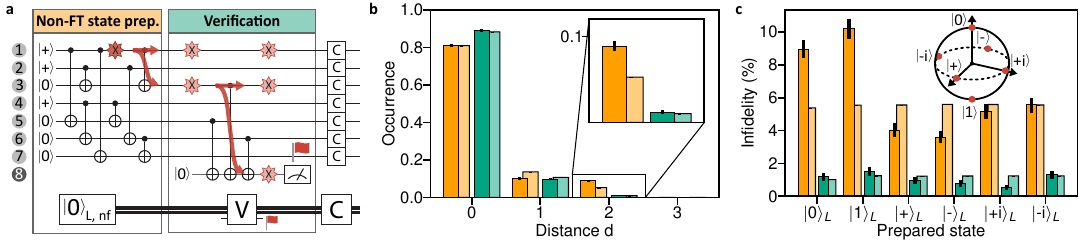}
    \caption{Fault-tolerant (FT) preparation of a logical basis state $\ket{0}_L$ and logical Clifford operations.
    (a) Logical Pauli states are prepared fault-tolerantly in three steps: First $\ket{0}_L$ is prepared by a non-fault-tolerant circuit. fault tolerance is ensured through verification (V) of the state by coupling to an additional flag ancilla qubit. This qubit, when measured as $\ket{0}$, signals that the correct state has been prepared fault-tolerantly, i.e.~up to single-qubit errors. To prepare a logical Pauli eigenstate other than $\ket{0}_L$ an additional transversal Clifford gate needs to be applied.
    (b) Relative occurrence rates of logical output states of distance $d$ to the target state $\ket{0}_L$ for non-FT (orange) and FT (turquoise) initialization. Example states of $d=1,2,3$ are $X_0\ket{0}_L, X_0X_1\ket{0}_L,X_0X_1X_2\ket{0}_L$. Simulation results are depicted by lighter colored bars. as described in the main text, all circuit elements are subject to depolarizing noise in numerical simulations.
    (c) Logical infidelities of all six logical Pauli eigenstates (red markers on Bloch sphere) including an ideal round of error correction performed in post-processing (experimental/simulation results depicted darker/lighter).
    }
    \label{fig:FT_prep}
\end{figure*}

We start by experimentally preparing the logical state $\ket{0}_L$ as the $+1$-eigenstate of the logical Z-operator $Z_L$ by implementing the circuit shown in Fig.~\ref{fig:FT_prep}a. The first part of the circuit encodes a logical qubit in a non-FT fashion: A single error, e.g.~a bit flip on qubit 1, can propagate to other qubits, here qubit 3, resulting in an uncorrectable weight-2 error and therefore causing a logical error. To render this encoding circuit fault-tolerant, a verification step is added (see Fig.~\ref{fig:FT_prep}a)~\cite{Goto2016, Bermudez2019}. An additional ancilla qubit is used to herald a successful logical qubit initialization, meaning that for an ancilla measurement outcome of $+1$ no single error anywhere in the encoding circuit can have led to uncorrectable errors on the data qubit register. 
This \emph{flag} qubit will be triggered by any potentially detrimental error, as illustrated for the weight two error on qubits 1 and 3. For a measurement outcome of $-1$ of the flag qubit the initialization is aborted and repeated.

We analyze the quality of the information encoded in a logical qubit in terms of logical operator expectation values. After a projective measurement, the outcome of the logical operators is corrected according to the measured error syndrome, effectively performing a Pauli frame update~\cite{riesebos2017pauli, Knill2005} described in Sec.~\ref{sec:idealEC}.
The updated measurement outcomes are categorized by different error patterns: (i) The outcome of all three measured stabilizer generators is $+1$, meaning that the error syndrome is trivial, and the logical operator exhibits the expected outcome. (ii) After the correction suggested by the non-trivial syndrome, the expected logical state is recovered. This case corresponds to a single correctable error. (iii) The syndrome is again non-trivial, but after error correction the logical state is flipped, indicating an uncorrectable error. (iv) The logical state is flipped despite the syndrome being trivial, meaning a logical error occurred directly. These categories correspond to states that have a minimum Hamming distance to any constituent state of the logical zero state (see Sec.~\ref{sec:Pauli_states}, Eq.~\ref{eq:distance}) of 0, 1, 2 and 3, respectively. For outcomes in the categories (i) and (ii) the logical state is recoverable, whereas for (iii) and (iv) an uncorrectable logical error is induced. The relative occurrence of outcomes associated to those four categories for the initialization of $\ket{0}_L$ is shown in Fig.~\ref{fig:FT_prep}b. The verification circuit significantly suppresses the occurrence of errors leading to logical errors, resulting in decreased relative occurrences for error patterns (iii) and (iv) by a factor of more than 7.
A figure of merit describing the quality of the encoded state is the logical state fidelity, i.e. the overlap of the measured with the target logical Bloch vector. A detailed discussion of the logical fidelity, not to be confused with the standard quantum state fidelity, can be found in Sec.~\ref{sec:methods_theory}.
The logical infidelity is decreased from 0.090(4) to 0.012(1) by introducing the verification of the initialization, showing a clear signature that a FT implementation outperforms its non-FT counterpart despite the increased circuit complexity of one additional flag qubit and an increase from 8 to 11 entangling gates. The acceptance rate heralded by a $+1$ outcome of the flag measurement is 78.9(5)\%. This behavior is in good qualitative agreement with numerical simulations which yield infidelities of 0.0538(2) and 0.0101(1) for the non-FT and FT circuits respectively, and an acceptance rate of 84.42(4)\%.

\section{Transversal fault-tolerant operations}
\label{sec:transversal_gates}

\begin{figure}
\includegraphics[width=89mm]{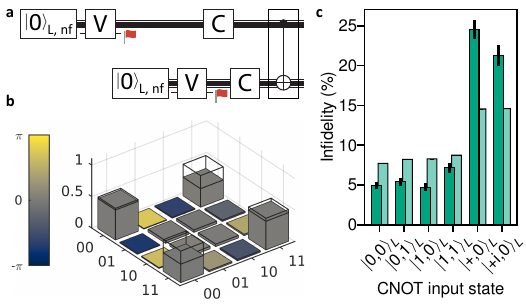}
\caption{FT implementation of a logical entangling gate. (a) To estimate the performance of the logical \mbox{CNOT-gate} we fault-tolerantly prepare six different logical two-qubit input states and apply the transversal \mbox{CNOT-gate} (framed gate at the end of the circuit).
(b) Logical state tomography after applying the \mbox{CNOT-gate} to the $\ket{+,0}$ state. The phase of the complex amplitudes is encoded in the color of the 3D bar plot and the wireframes depict ideal results.
(c) Logical infidelities for six different input states of the \mbox{CNOT-gate} (experimental/simulation results depicted darker/lighter).}
\label{fig:FT_cnot}
\end{figure}

The transversality of the Clifford group as a property of the color code allows for the preparation of the six cardinal states on the Bloch sphere, referred to as Pauli eigenstates herein, by applying single-qubit rotations corresponding to the respective logical gate to all qubits in the data register (see Fig.~\ref{fig:FT_concepts}b). For both experimental data but also results of numerical simulations, the verification of the initialization reduces the logical infidelity of all six Pauli eigenstates, as can be seen in Fig.~\ref{fig:FT_prep}c (see also Sec.~\ref{sec:methods_theory} for definition of the logical fidelity). Due to miscalibration of experimental parameters the non-FT preparation of $\ket{0}_L$ and $\ket{1}_L$ exhibit an increased infidelity, while the infidelity is comparable for all six states for the FT scheme. The imperfections introduced by the single-qubit rotations constituting the Clifford operation are negligible compared to those originating from the initialization of the logical qubit. The average logical infidelity for the FT circuit is 0.011(1) with an acceptance rate of 80.6(2)\%, while simulations suggest an average infidelity of 0.01203(4). The six stabilizer generators were measured projectively to verify the preparation of the correct encoded states for the Pauli states. The averaged expectation value of the X-(Z-)type stabilizers is 0.826(3) (0.760(3)) for the FT and 0.842(3) (0.790(3)) for the non-FT preparation scheme. Further details on the measured stabilizer generators can be found in Sec.~\ref{sec:Pauli_states}.

The experimental implementation of the \mbox{CNOT-gate} acting on two logical qubits encoded in the color code is noticeably more challenging in terms of register size and circuit complexity, requiring 29 entangling gates applied to 16 qubits. As shown in Fig.~\ref{fig:FT_cnot}, we first prepare the two logical $\ket{0}_L$ states, then apply single-qubit rotations to prepare six different logical Pauli input states. The transversal logical \mbox{CNOT-gate} is implemented by sequentially applying \mbox{CNOT-gates} to corresponding pairs of physical qubits of the two logical qubits (see also Fig.~\ref{fig:FT_concepts}b). A single error on any of the physical qubits propagates to at most one error on each of the logical qubits and therefore remains correctable, thereby ensuring fault tolerance of the gate realization.
Applying the logical \mbox{CNOT-gate} to the input state $\ket{+,0}_L$ yields the logical Bell state $\frac{1}{\sqrt{2}}(\ket{0,0}_L + \ket{1,1}_L)$ depicted in Fig.~\ref{fig:FT_cnot}b. This logical density matrix is retrieved from measurements of all nine twofold combinations of the logical observables $X_L$, $Y_L$ and $Z_L$ on the two logical qubits, followed by a maximum likelihood state reconstruction of the two-qubit state~\cite{Hradil2004}.
The logical fidelity of the logical Bell state is 0.754(9), consequently the correlations of the logical operator outcomes correspond to those of a Bell state with clear two-partite entanglement.
Figure~\ref{fig:FT_cnot}c shows the logical infidelity for six different input states. It reveals that the infidelity of output states is higher if the control qubit is in a superposition state, thus leading to an entangled outcome, compared to cases in which the outcome is a basis state of the logical two-qubit computational basis. This increased error rate is well-described by numerical simulations based on the circuit noise model: The average logical infidelity is $0.110^{+3}_{-4}$ and $0.1035(1)$ for the experimental implementation and the numerical simulation, respectively.

\section{Universal fault-tolerant operations}
\label{sec:universal_gates}

\begin{figure*}
    \centering
    \includegraphics[width=183mm]{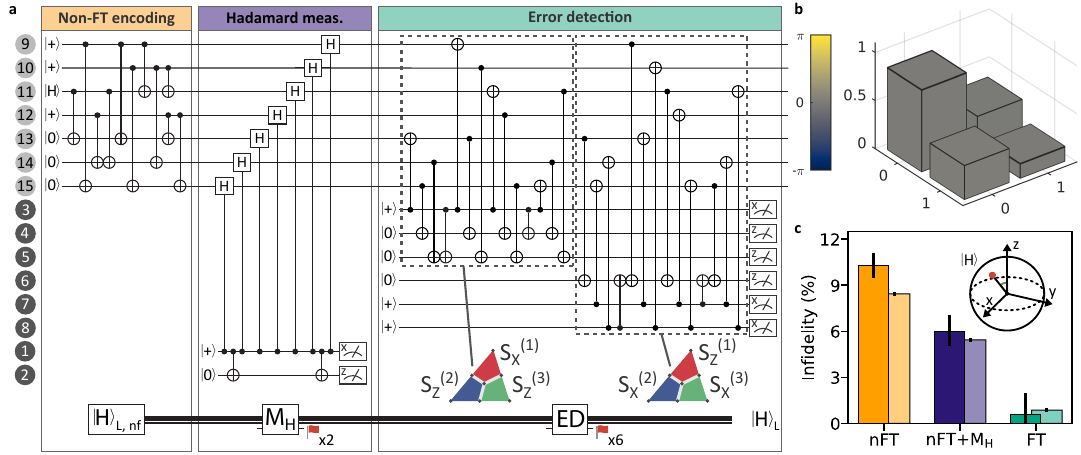}
    \caption{Fault-tolerant generation of a logical magic state $\ket{H}_L$ (see Eqn.~{\ref{eq:magicstate}}).
    (a) The magic state is prepared non-fault-tolerantly in a first step, where a physical magic state $\ket{H}$ is mapped to the logical state $\ket{H}_{L,\text{nf}}$ encoded in the data qubits at positions 9 to 15 in the ion string (see labels at left of circuit). Thereafter, a FT measurement of the Hadamard operator ($M_H$) is carried out. Two ancilla qubits herald that the prepared state is a $+1$-eigenstate of the Hadamard operator but also that no dangerous error occurred during the measurement. The magic state preparation is concluded with an error detection block that measures the three X- and Z-type stabilizers each in a fault-tolerant fashion. The first part of the error detection circuit (first dashed box), measures $S_X^{(1)}$, $S_Z^{(2)}$ and $S_Z^{(3)}$, whereas the second part measures $S_Z^{(1}$, $S_X^{(2)}$ and $S_X^{(2)}$. The magic state preparation is discarded and repeated in case of a non-trivial syndrome of the eight ancilla qubits 1 to 8.
    (b) Logical state tomography (see Sec.~\ref{sec:transversal_gates}) after FT magic state preparation. The phase of the complex amplitudes is encoded in the color of the 3D bar plot and the wireframes depict ideal results. Phase deviations from the ideal density matrix are smaller than $\SI{50}{\milli\radian}$ while amplitude deviations are smaller than $0.07$.
    (c) The decrease in infidelity of the logical magic state (red marker on Bloch sphere) after each step of the FT preparation procedure is observed experimentally and captured by depolarizing noise simulations (experimental/simulation results depicted darker/lighter).}
    \label{fig:FT_t_gen}
\end{figure*}

\begin{figure}
    \centering
    \includegraphics[width=89mm]{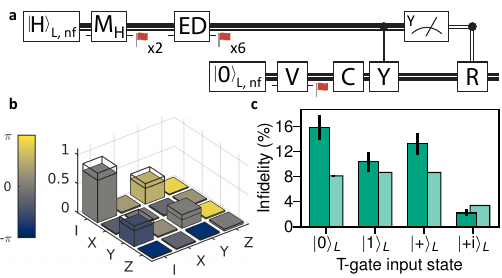}
    \caption{FT \mbox{T-gate} injection. (a) After performing the fully fault-tolerant three step procedure of preparing the logical magic state, the logical \mbox{T-gate} is applied via logical gate teleportation onto a second register that has a logical Pauli state prepared. Conditional application of $R \equiv R_Y(\pi/2)$ is done in post-processing. (b) Logical process matrix of the experimental logical \mbox{T-gate}. The phase of the complex amplitudes is encoded in the color of the 3D bar plot and the wireframes depict ideal results. (c) Infidelities of the data qubit state when applying the logical \mbox{T-gate} to several logical Pauli input states (experimental/simulation results depicted darker/lighter). Infidelity is the lowest for the $\ket{+i}_L$ state since it is an eigenstate of the \mbox{T-gate}. Infidelity is slightly larger for $\ket{1}_L$ and $\ket{+}_L$ than for $\ket{0}_L$ since preparation contains an additional transversal Clifford operation prone to errors.}
    \label{fig:FT_ti_inj} 
\end{figure}

A universal set of logical gates allows to implement any logical unitary operation to arbitrary precision. The ability to perform a $\pi/4$-rotation about any axis is known to be sufficient to augment the set of Clifford gates, which are transversal in the color code, to a universal gate set.
The logical T-gate 
\begin{align}
    T_L = e^{-i\frac{\pi}{8}Y_L}
\end{align}
performs a $\pi/4$-rotation about the Y-axis and can be implemented by magic state injection as shown in Fig.~\ref{fig:FT_concepts}b. It consists of the logical CNOT-operation we have demonstrated in the preceding section, a logical measurement and single-qubit Clifford operation conditioned on the logical measurement outcome. First preparing and then injecting the logical magic state 
\begin{align}
    \ket{H}_L &= \cos(\pi/8)\ket{0}_L + \sin(\pi/8)\ket{1}_L
    \label{eq:magicstate}
\end{align}
enables gate teleportation of the logical non-Clifford \mbox{T-gate}. The logical magic state in Eq. (\ref{eq:magicstate}) is the $+1$-eigenstate of the logical Hadamard operator. Its corresponding $-1$-eigenstate is $\ket{-H}_L = Y_L\ket{H}_L$, related to $\ket{H}_L$ via a logical Y-flip. 

Recently, a resource-efficient procedure to prepare a magic state using FT circuits following the flag fault tolerance paradigm has been proposed \cite{Chamberland2019}. The procedure consists of the following steps, depicted in Fig.~\ref{fig:FT_t_gen}a: we begin with a non-FT preparation of the magic state $\ket{H}_{L,\text{nf}}$, as recently demonstrated also in \cite{Ryan-Anderson2021} (shown in subbox "Non-FT encoding"). Next, a measurement of the logical Hadamard operator is performed, which projects input states onto the $+1$-eigenspace and discards states that are eigenstates with eigenvalue $-1$ (subbox "Hadamard meas."). The latter may be caused by single faults in the circuit (e.g.~faults in the initial state preparation of $\ket{H}$ for physical qubit 11), thus rendering the circuit non-FT if $-1$-eigenstates are not discarded in this step. Here both ancillas (qubits 1 and 2) are utilized as flag qubits. The syndrome measurement ancilla flags when the $-1$-eigenstate has erroneously been prepared, the second ancilla flags when a dangerous fault has occurred that may corrupt the state. The last step is a complete error correction (EC) cycle, consisting of fault-tolerantly measuring all six stabilizers of the color code using one flag qubit per stabilizer as suggested in \cite{Chao2018}. The EC block is used to sort out faulty states whenever any flag qubit is measured as $-1$ and thus enables error detection (ED) (subbox "Error detection"). The resulting states after performing all three steps are guaranteed to be the correct logical magic state $\ket{H}_L$ up to correctable single-qubit errors provided at most one fault has occurred anywhere in the circuit. 
In total the protocol requires eight ancilla qubits that act as flags. The generated state is accepted as valid if all flag qubits indicate that no harmful error has happened.

We implement all steps for the magic state preparation and estimate the logical infidelities of the generated magic state. We compare the results to numerical simulations of depolarizing noise on all circuit operations. In Fig.~\ref{fig:FT_t_gen}b the reconstructed density matrix of the fault-tolerantly prepared magic state is shown and its ideal numerical values can be found in Eq.~(\ref{eq:rho_H}).
Fig.~\ref{fig:FT_t_gen}c shows the logical infidelity of the magic state, which clearly decreases after each preparation step in both experiment and simulation. Each step of the magic state initialization process improves the quality of the generated logical state.
After the full FT initialization procedure, a logical infidelity of $0.006^{+14}_{-5}$ for the magic state $\ket{H}_L$ with an acceptance rate of 13.7(3)\% is found in the experimental realization, whereas numerical simulations predict approximately 27\%, see Sec.~\ref{sec:acceptance_rate} for discussion.

Next, the fault-tolerant magic state initialization is followed by transversal Clifford operations to fault-tolerantly teleport the logical magic state, thereby resulting in a realization of a FT logical \mbox{T-gate}. For this, we perform an in-sequence measurement of the flag qubits for the magic state generation as sketched in Fig.~\ref{fig:FT_ti_inj}a. In the case of heralded successful magic state generation, the ancilla qubits are in a well-defined state after the measurement and can directly be re-used to encode a second logical qubit in $\ket{0}_L$ using the FT protocol from Fig.~\ref{fig:FT_prep}a. We then apply a transversal Clifford operation on this second logical qubit to prepare one of the logical initial states $\ket{0}_L,\,\ket{1}_L,\,\ket{+}_L,\,\ket{+i}_L$. Finally, the transversal controlled-Y operation is applied on the second register and all physical qubits are measured. The measurement outcome for the logical Y-operator of the control qubit in the first register is then extracted and the conditional Y$(\pi/2)$-rotation $R$ to the target qubit in the second register is applied in post-processing, see Sec.~\ref{sec:methods_theory} for details. By measuring the logical state of the target register for the four different initial states, it is possible to reconstruct the logical process matrix, shown in Fig.~\ref{fig:FT_ti_inj}b with the ideal values explained in Sec.~\ref{sec:process_matrix}.
Fig.~\ref{fig:FT_ti_inj}c shows the logical infidelities for the different input states, yielding a mean infidelity of 0.10(1).
It is expected and indeed observed experimentally that the best fidelity is achieved for the logical $Y$-eigenstate $\ket{+i}_L$ as it is an eigenstate of the \mbox{T-gate}. Infidelities for the three other logical input states are slightly higher, which qualitatively agrees with the numerical simulations.

\section{Discussion and Outlook}

In this work we have demonstrated the first fault-tolerant implementation of a universal set of single- and two-qubit logical gates. We were able to witness a hallmark feature of fault-tolerant circuit design, namely an improvement of the performance of encoded qubits, despite the FT implementations of encoding and manipulation requiring an increased gate count and complexity of the underlying circuits. The resource-efficient implementation of these FT operations is enabled by the all-to-all qubit connectivity in the present trapped ion architecture, allowing for entangling operations between arbitrary pairs of qubits.
Predictions from numerical simulations based on a relatively simple, generic and architecture-agnostic depolarizing circuit noise model, only informed by estimated experimental error rates, approximate the experimental findings well. The largest deviations between experimental behavior and and numerical predictions were observed for the logical \mbox{CNOT-gate}. A more extensive characterization of this logical entangling gate and the other fault-tolerant gadgets,  together with more sophisticated and validated theoretical noise models will be subject to future investigations, and is imperative for designing future QEC architectures and procedures.

On the way towards error-protected universal quantum computation on even more robust logical qubits, further milestones ahead are the incorporation of repetitive QEC cycles~\cite{Ryan-Anderson2021} into the FT logical gate operations demonstrated in our work. Another hurdle to be taken is the demonstration of error correction and FT gate operations for larger-distance logical qubits~\cite{Chamberland2019}.

\clearpage

\section{Experimental methods}
\label{sec:methods_exp}

The experiments described in this work are performed on a trapped ion quantum computer. $^{40}$Ca$^{+}$ ions are trapped in a macroscopic Paul trap and the optical qubit is encoded in two Zeeman sublevels of the 4$S_{\nicefrac{1}{2}}$ and 3D$_{\nicefrac{5}{2}}$ electronic states. Further details on the experimental setup can be found in the recent publication~\cite{Pogorelov2021}.

\subsection{Trapping and cooling}
For this work, the ion crystal is configured to consist of 16 ions with an axial center-of-mass (COM) mode frequency $\omega_\textrm{ax, COM} = 2\pi \times \SI{400}{\kilo\Hz}$  and radial COM mode frequencies of $\omega_{\textrm{rad1,COM}} = 2\pi \times \SI{3270}{\kilo\Hz}$ and $\omega_{\textrm{rad2,COM}} = 2\pi \times \SI{3100}{\kilo\Hz}$.
Before executing any gate sequence, the radial motional modes of the ion chain are cooled nearly to the ground state via Doppler cooling for \SI{2}{\milli\s}, followed by resolved sideband cooling for \SI{15}{\milli\s}. Subsequently, the qubits are initialized via optical pumping to the 4$S_{\nicefrac{1}{2}, m_j = -\nicefrac{1}{2}}$ ground state.

\subsection{Qubit manipulation}
Coherent qubit manipulation is performed by individually addressable laser pulses at a wavelength of \SI{729}{\nano\m}. Pulses resonant with the 4$S_{\nicefrac{1}{2}, m_j = -\nicefrac{1}{2}}$ to 3$D_{\nicefrac{5}{2}, m_j = -\nicefrac{1}{2}}$ transition enable rotations around an arbitrary axis in the equatorial plane of the Bloch sphere, where the angle between the rotation axis and the X-axis is determined by the phase $\phi$ of the light pulse. Those operations are described by $R_{\phi}^{(i)}(\theta) = \mathrm{exp}(-i\frac{\theta}{2} (\sigma_x^{(i)} \cos{\phi} - \sigma_y^{(i)} \sin{\phi}))$, where $\sigma_x^{(i)}$ and $\sigma_y^{(i)}$ are single-qubit Pauli matrices acting on qubit $i$. Rotations around the X-axis $R_x$ (Y-axis $R_y$) can be implemented by setting $\phi$ to $0$ ($-\pi/2$). A pulse length of \SI{15}{\micro\s} is required to implement a $\pi/2$-pulse on a single qubit. Randomized benchmarking for single-qubit gates in the 16-ion chain yields an average fidelity of a $\pi/2$-gate of $99.51 \pm 0.05 \%$. Additionally, rotations around the Z-axis of the Bloch sphere for a specific ion can be implemented \emph{virtually} by introducing a phase shift to all subsequent pulses applied to the ion.

\begin{figure}
    \centering
    \includegraphics[width=\columnwidth]{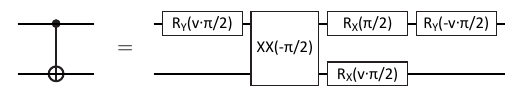}
    \caption{Decomposition of a \mbox{CNOT-gate} into gates native to the experimental architecture used in this work. The parameter $v = \pm 1$ can be varied arbitrarily.}
    \label{fig:cnot_decomp}
\end{figure}

Two-qubit gates are realized by the \MS interaction \cite{Sorensen2000} described by $\mathrm{MS}_{ij}(\theta) = \mathrm{exp}(-i\frac{\theta}{2} \sigma_x^{(i)}\sigma_x^{(j)})$. An arbitrary pair of ions is addressed with bichromatic beams slightly detuned from the radial COM mode $\omega_{\textrm{rad1,COM}}$. Gate time $t_\text{gate} = \SI{270}{\micro\s}$ and detuning from the COM mode $\Delta\approx 2\pi \times \SI{3.7}{\kilo\Hz}$ are chosen to allow for simultaneous decoupling of the two closest radial modes $\omega_{\textrm{rad1,COM}},~ \omega_{\textrm{rad1,2}}$ at the end of the interaction.
An additional (third) frequency tone, \SI{1.05}{\mega\Hz} blue-detuned from the carrier transition, compensates the AC Stark shift induced by the bichromatic light field.
For $\theta = \pi/2$ this results in an \mbox{XX-gate} which is equivalent to a \mbox{CNOT-gate} up to single-qubit rotations~\cite{Maslov2017}, as depicted in Fig.~\ref{fig:cnot_decomp}. The average Bell state fidelity in a chain of 16 ions is about 97.5\% for entangling gates between neighbouring ions.

\subsection{State readout}
Qubit state readout is performed by illuminating the ions with a light field resonant to the  4$S_{\nicefrac{1}{2}}$ to 4$P_{\nicefrac{1}{2}}$ transition and collect scattered photons. Due to technical limitations imposed by the EMCCD camera, site-resolved state readout is only possible after the coherent evolution. In-sequence detection events utilizing an avalanche photodiode (APD) can only reveal the number of excitations present in the ion string. 
A subset of qubits can be read out in-sequence by shelving the population in the 4$S$ state of all other ions to the 3$D_{\nicefrac{5}{2}, m_j = -\nicefrac{3}{2}}$ Zeeman sublevel prior to the illumination of the ion string with the detection light field. This technique is used for the measurements presented in Sec.~\ref{sec:universal_gates}. For the FT initialization of the magic state the ancilla qubits are measured via the APD. If there are no excited ions detected, the protocol is continued by reusing the measured ions for encoding a second logical qubit state and injecting the magic state.
After an illumination time of \SI{2}{\milli\s} for the EMCCD measurement and \SI{0.5}{\milli\s} for the APD measurement a readout fidelity of $>99.7$\% is achieved, where this number refers to the single-qubit readout fidelity for EMCCD measurements and and the discrimination between $0$ and $>0$ excited qubits for APD measurements.

\subsection{Error estimation}
The errors given throughout this work solely account for statistical errors. For the estimation of the statistical fluctuations all measured outcomes are resampled from a multinomial distribution according to their respective probabilities. The stated errors in the text but also errorbars given in figures correspond to 68\% confidence intervals extracted from the resampled datasets.

\begin{figure*}
\includegraphics[height=210mm]{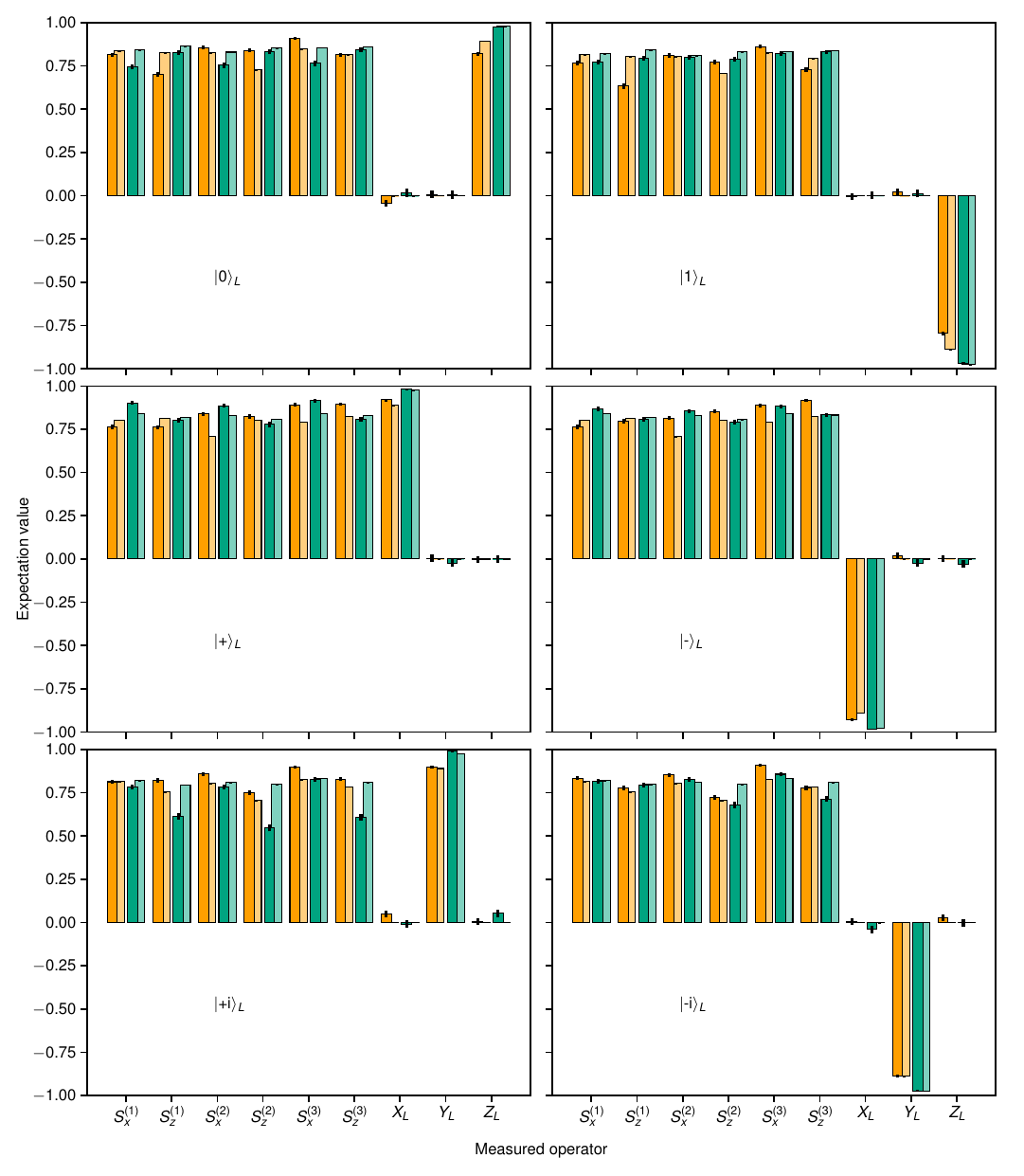}
\caption{Expectation values of the stabilizer generators and the logical operators of the seven-qubit color code for the six cardinal states of the Bloch sphere. Results for the non-FT and FT preparation scheme are depicted in orange and turquoise respectively, whereas results from numerical simulations are shown in lighter colored bars. $2500$ and $10^6$ runs were performed in the experiment and for simulations for each prepared state, respectively. For the calculation of the expectation values of the logical operators a round of perfect error correction is applied. For the measurements corresponding to the data presented in this figure but also in Fig.~\ref{fig:FT_prep} the sign of the rotation angle of physical Y-rotations is flipped, effectively implementing an additional deterministic $\pi$ phase flip on qubit 6 and a $\pi$ bit flip on qubit 7 at the end of the circuit depicted in Fig.~\ref{fig:FT_prep}a. The effects of this redefinition do not amount in a change of measurement bases and can be readily accounted for in post-processing.}
\label{fig:stabilizers}
\end{figure*}

\section{Simulation methods}
\label{sec:methods_theory}
Theoretical simulation results presented in the main text are obtained using stabilizer simulations and state\-vector simulations for the logical Pauli states and magic state preparation and injection circuits, respectively. We use the "Performance Estimator of Codes On Surfaces" (PECOS) package due to its flexibility in analyzing error propagation in different error models through Monte Carlo (MC) simulation (publicly available at \mbox{\url{https://github.com/PECOS-packages/PECOS}}) \citep{ryan2018quantum}. In these simulations any ideal circuit element is replaced by a faulty element, consisting of the ideal operation followed by an error operator, with a given probability.

We model circuit errors as depolarizing errors, which reproduces well the experimentally observed infidelities despite its conceptual simplicity which does not take the microscopic physical processes underlying noisy gates and operations in the ion trap into account explicitly. Noise is applied in simulations by randomly placing Pauli errors $E$ according to the experimental physical error rates after every single-qubit operation, i.e. single-qubit gates, initializations and measurements, as well as two-qubit gates, each with their respective error rates $p_1,\,p_i,\,p_m$ and $p_2$. These errors can be
\begin{align}
	E_x &\in \{\sigma_k, \forall k \in \{1,2,3\}\},~x \in \{1,i,m\} \\
	E_2 &\in \{\sigma_k \otimes \sigma_l, \forall k,l \in \{0,1,2,3\}\}~\backslash~\{I \otimes I\}
\end{align}
where $\sigma_k = \{I,X,Y,Z\}$ with $k=0,1,2,3$ are the Pauli matrices.
The error channels for our depolarizing noise model read
\begin{align}
    \mathcal{E}_x(\rho) &= (1-p_x)\rho + \frac{p_x}{3}(X\rho X + Y\rho Y + Z \rho Z) \\
    \mathcal{E}_2(\rho) &= (1-p_2)\rho + \frac{p_2}{15} \left[ \sum_{i,j=0}^3 (\sigma_i \rho \sigma_j) - \rho \right]
\end{align}
so that any single-qubit error is applied uniformly to the ideal operation with equal probability $p_x/3$ and the single-qubit operation is executed ideally with probability \mbox{$1-p_x$}; two-qubit errors are applied uniformly after the ideal two-qubit gates with equal probability $p_2/15$ and any two-qubit gate is executed ideally with probability $1-p_2$.
In all simulations we used physical error rates of
\begin{align}
    p_1 &= 0.005 \notag \\
    p_2 &= 0.025 \label{eq:MC_errorrates} \\
    p_i &= p_m = 0.003 \notag
\end{align}
for the corresponding operations.

The logical Pauli state encoding circuits (Fig. \ref{fig:FT_prep}a) and logical CNOT circuit (Fig. \ref{fig:FT_cnot}a) are simulated efficiently using stabilizer simulations. This is possible since we are preparing eigenstates of Pauli operators, measuring exclusively in one of three possible Pauli bases here and these circuits only contain Clifford gates.

This description as stabilizer states breaks down when arbitrary single-qubit rotations are to be performed by the circuit, especially with regards to the magic state, Eq.~(\ref{eq:magicstate}), that generates the non-Clifford T-gate.
The circuits that fault-tolerantly prepare the logical magic state (Fig.~\ref{fig:FT_t_gen}a) and perform the gate teleportation (Fig.~\ref{fig:FT_ti_inj}a) contain non-Clifford operations and thus we are required to run full statevector simulations. 

In the teleportation circuit (Fig.~\ref{fig:FT_ti_inj}a), the logical controlled-Y is followed by measurement of all data qubits of the first register in the Y-basis and  application of a classically controlled Y-rotation
\begin{align}
    R \equiv R_Y(\pi/2)
\end{align}
depending on the measurement result of the first register where the logical magic state has been prepared previously. The logical gate $R_L \simeq R^{\otimes 7}$ is applied to the second register which carries a logical Pauli state, e.g. $\ket{+}_L$. The resulting output state is the logical T-gate applied to the logical Pauli state, e.g. $T_L\ket{+}_L$. Both in simulation and experiment the effect of the R-gate is taken into account by altering the destructive final data qubit measurements. Since $R$ is a $\pi/2$-rotation about the Y-axis, it maps Z-basis states onto X-basis states and vice versa.

\section{Characterization methods}
\label{sec:methods_char}

\subsection{Ideal error correction}
\label{sec:idealEC}
Whenever performing destructive measurements on encoded data qubits we may reinterpret the measurement result according to the color code look-up table decoder. For example, from measuring the bitstring $0000001$ on a seven-qubit register we may conclude for low physical error rates present in our setup that the likeliest error on those qubits has been a single X-flip on the first qubit and reinterpret the measurement result as $0000000$. This process of ideal or in-software error correction (EC) is commonly used and possible whenever one aims not to keep running further quantum circuits on the error-corrected state. In general the corrected bitstring is determined by extracting the syndrome from the overlap of the measured bitstring with the stabilizer generators in binary notation and applying the respective correction. For CSS codes such as the color code the X- and Z-sectors can be treated distinctly. So when measuring the bitstring $0000001$ in the Z-basis, the overlap with Z-stabilizers in binary notation $s_Z^{(1)} = 1010101, s_Z^{(2)} = 1111000, s_Z^{(3)} = 1100110$ yields the syndrome $[-1,+1,+1]$. Here binary 1s correspond to a Pauli Z-operator for the single qubit at the respective position and binary 0s represent the identity operation. Reinterpreting the measured bitstring as $0000000$ is equivalent to applying a $X_1$ correction operator based on the syndrome information which would correctly recover the original state. Since in this work we are demonstrating FT operations, all final quantum states may only be correct up to an arbitrary single Pauli error. These errors are accounted for via ideal error correction.

\subsection{Logical Pauli states}
\label{sec:Pauli_states}

When fault-tolerantly encoding logical Pauli states we characterize the output state by categories of errors present after executing the circuit in Fig.~\ref{fig:FT_prep}a. The categories of errors given in Fig.~\ref{fig:FT_prep}b refer to the states of distance $d$ to the desired $\ket{0}_L$-state which determines whether or not we can correctly identify the state as $\ket{0}_L$ after ideal EC. 

We obtain the distance $d$ to the desired $\ket{0}_L$-state by destructively measuring the data qubit register and finding the minimal Hamming distance $D_H$ of the measurement bitstring $m$ to the bitstrings that label the basis states of
\begin{align}
    \ket{0}_L = \frac{1}{\sqrt{8}}\Big(\ket{0000000} &+ \ket{1010101} + \ket{0110011} \notag \\ 
    + \ket{1100110} &+ \ket{0001111} + \ket{1011010} \notag \\
    + \ket{0111100} &+ \ket{1101001}\Big) \label{eq:steane_zero}
\end{align}
in post-processing:
\begin{align}
    d \equiv \min\Big( D_H(m,0000000), D_H(m,1010101)&, \notag \\ \dots , D_H(m,1101001) \Big) &\text{.} \label{eq:distance}
\end{align}

Ideal EC will trivially correct the exact $\ket{0}_L$ state, i.e. $d=0$, and correct all states with single Pauli errors, e.g. $X_2 \ket{0}_L$ ($d=1$).
It will yield the logically flipped $\ket{1}_L$ result when acting on a state of distance $d=2$ or $d=3$ from $\ket{0}_L$, i.e. $\ket{0}_L$ carrying two X-errors or directly a weight-3 logical bit flip $X_L$.

For arbitrary logical Pauli states destructive measurements must be performed in their respective basis. However, only Pauli-X (Pauli-Z) errors are visible in the preparation of logical Pauli-Z (Pauli-X) basis states $\ket{0}_1,\,\ket{1}_L\,(\ket{\pm}_L)$.

Note that for all measurements on the characterization of logical Pauli states shown in Fig.~\ref{fig:FT_prep} and Fig.~\ref{fig:stabilizers} an accidental redefinition of the rotation direction of physical single-qubit Y-rotations is accounted for in post-processing. See Fig.~\ref{fig:stabilizers} for more details.

\subsection{Logical fidelities}

\textbf{Single-qubit logical states.} --
The logical fidelities presented in the main text are obtained by reconstructing the logical Bloch vector of the prepared state $\rho$ and determining the overlap with the Bloch vector of a logical target state. Within the code space, the projector onto an ideal single-qubit logical target state $\rho_t = \ket{t}\bra{t}_L$ is given by 
\begin{align}
    P_t &= \frac{1}{2}\left( I + O_t \right)
\end{align}
with $O_t$ the logical operator that the target state $O_t\ket{t}_L = \ket{t}_L$ is the $+1$-eigenstate to. For the Pauli states considered in this work the projectors are
\begin{align}
    P_{0/1} &= \frac{1}{2}\left( I \pm Z_L \right) \label{eq:proj0} \\
    P_{\pm} &= \frac{1}{2}\left( I \pm X_L \right) \\
    P_{\pm i} &= \frac{1}{2}\left( I \pm Y_L \right) \label{eq:projI}
\end{align}
and logical fidelity of a prepared state $\rho$ follows as
\begin{align}
    \mathcal{F}_t(\rho) &= \langle P_t \rangle = \tr \left( P_t \rho \right) \text{.} \label{eq:logfid}
\end{align}
We emphasize that these logical fidelites are \emph{not} equivalent to the full quantum state fidelities $\mathcal{F} = \tr \left( \rho_t \rho \right)$ but are the probabilities to be able to correctly conclude which logical state was intended to be prepared or stored. 

Combining Eqs.~(\ref{eq:proj0})-(\ref{eq:projI}) with the expression for the logical fidelity in Eq.~(\ref{eq:logfid}), we can see that expectation values of logical Pauli operators $O_t$ need to be determined in order to find the logical fidelities 
\begin{align}
    \mathcal{F} &= \frac{1}{2}\left( 1 \pm \langle O_t \rangle \right) ~\text{with}~O_t \in \{X_L,Y_L,Z_L\}, \label{eq:Pauli_fid}
\end{align}
which e.g.~evaluates to $\mathcal{F}_{0} = \frac{1}{2}\left( 1 + \langle Z_L \rangle \right)$ for the logical Pauli state $\ket{0}_L$. All six cardinal state logical fidelities are shown in Fig.~\ref{fig:FT_prep}c. We sample the expectation values of the logical Pauli operators by running stabilizer simulations of the respective preparation circuit $N=10^6$ times followed by destructive measurement of all data qubits and ideal EC in the respective Pauli basis. The measurement result for a logical operator before EC is determined as $(-1)^\abs{m}$ by the number of 1s in the measurement bitstring $m$ modulo 2. Then a round of ideal EC as described in Sec.~\ref{sec:idealEC} is performed to obtain the final measurement result. Measurement results from each run are averaged to obtain the expectation value of the respective logical operator.   

\textbf{Two-qubit logical states.} --
In order to characterize the logical \mbox{CNOT-gate} through stabilizer simulations we first list its action on various logical input states in the table below: 
\begin{center}
\begin{tabular}{l|l}
input & output \\ \hline 
$\ket{0,0}_L$ & $\ket{0,0}_L$ \\
$\ket{0,1}_L$ & $\ket{0,1}_L$ \\
$\ket{1,0}_L$ & $\ket{1,1}_L$ \\
$\ket{1,1}_L$ & $\ket{1,0}_L$ \\
$\ket{+,0}_L$ & $\ket{\beta}_L$ \\
$\ket{+i,0}_L$ & $\ket{\gamma}_L$ 
\end{tabular}
\end{center}
where $\ket{x,y}_L = \ket{x}_L \otimes \ket{y}_L$. Single logical qubit states $\ket{x}_L$ and $\ket{y}_L$ are prepared distinctly in two seven-qubit registers. The \mbox{CNOT-gate}, acting $\ket{x,y}_L \mapsto \ket{x,y\oplus x}_L$, flips the second bit (target) if the first bit (control) is in the 1-state. Thus, $\ket{+,0}_L$ is mapped to the maximally entangled Bell state 
\begin{align}
    \ket{\beta}_L = \frac{1}{\sqrt{2}}(\ket{0,0}_L+\ket{1,1}_L), 
\end{align}
which can equivalently be expressed by the logical density operator
\begin{align}
    \rho_\beta &= \ket{\beta}\bra{\beta}_L \simeq \frac{1}{2} \begin{pmatrix}
    1 & 0 & 0 & 1\\
    0 & 0 & 0 & 0\\
    0 & 0 & 0 & 0\\
    1 & 0 & 0 & 1
    \end{pmatrix} \label{eq:rho_Bell}
\end{align}
where the matrix representation is in the logical two-qubit computational basis. Quantum state tomography has been performed to quantify experimental capabilities to obtain the logical Bell state described by this density matrix as shown in Fig. \ref{fig:FT_cnot}b. Expectation values of all logical two-qubit Pauli matrices including the identity are measured and subsequently maximum likelihood techniques are used to reconstruct the logical density operator~\cite{Hradil2004}. The Bell state is stabilized by the logical operators $Z^1_LZ^2_L$ and $X^1_LX^2_L$ where superscripts $.^{1,2}$ refer to the two logical qubits.
Analogously, the CNOT maps input $\ket{+i,0}_L$ to the Y-basis maximally entangled state 
\begin{align}
    \ket{\gamma}_L = \frac{1}{\sqrt{2}}(\ket{0,0}_L+i\ket{1,1}_L).
\end{align} 
Its stabilizers can be obtained by realizing that both states are related via a phase gate 
\begin{align}
    \ket{\gamma} = S_1\ket{\beta}
\end{align} so by transforming the stabilizer generators of $\ket{\beta}$ as
\begin{align}
    S_1 Z_1Z_2 S_1^\dagger &= Z_1Z_2 \\
    S_1 X_1X_2 S_1^\dagger &= X_1Y_2 
\end{align}
we obtain the stabilizer generators of $\ket{\gamma}$.

The projectors onto the logical two-qubit output states we wish to characterize is now given by the product of the projectors onto the simultaneous $+1$-eigenspace of the logical operators in both registers
\begin{align}
    P_{00} &= (P_0 \otimes I)(I \otimes P_0) = \frac{1}{2}\left( I + Z_L^1 \right)\frac{1}{2}\left( I + Z_L^2 \right) \\
    P_{01} &= (P_0 \otimes I)(I \otimes P_1) = \frac{1}{2}\left( I + Z_L^1 \right)\frac{1}{2}\left( I - Z_L^2 \right) \\
    P_{11} &= (P_1 \otimes I)(I \otimes P_1) = \frac{1}{2}\left( I - Z_L^1 \right)\frac{1}{2}\left( I - Z_L^2 \right) \\
    P_{10} &= (P_1 \otimes I)(I \otimes P_0) =\frac{1}{2}\left( I - Z_L^1 \right)\frac{1}{2}\left( I + Z_L^2 \right) \\
    P_\beta &= \frac{1}{2}\left( I + X^1_LX^2_L \right) \frac{1}{2}\left( I + Z^1_LZ^2_L \right) \\
    P_\gamma &= \frac{1}{2}\left( I + Z^1_LZ^2_L \right) \frac{1}{2}\left( I + X^1_LY^2_L \right)
\end{align}

Employing Eq.~(\ref{eq:logfid}), the logical fidelities for the output states of the logical \mbox{CNOT-gate} follow as expectation values of the logical two-qubit state projectors as
\begin{align}
    \mathcal{F}_{00} &= \frac{1}{4}\left( 1 + \langle Z_L^1 \rangle + \langle Z_L^2 \rangle + \langle Z_L^1Z_L^2 \rangle \right) \\
    \mathcal{F}_{01} &= \frac{1}{4}\left( 1 + \langle Z_L^1 \rangle - \langle Z_L^2 \rangle - \langle Z_L^1Z_L^2 \rangle \right) \\
    \mathcal{F}_{11} &= \frac{1}{4}\left( 1 - \langle Z_L^1 \rangle - \langle Z_L^2 \rangle + \langle Z_L^1Z_L^2 \rangle \right) \\
    \mathcal{F}_{10} &= \frac{1}{4}\left( 1 - \langle Z_L^1 \rangle + \langle Z_L^2 \rangle - \langle Z_L^1Z_L^2 \rangle \right) \\
    \mathcal{F}_\beta &= \frac{1}{4}\left( 1 + \langle X_L^1X_L^2 \rangle - \langle Y_L^1Y_L^2 \rangle + \langle Z_L^1Z_L^2 \rangle \right) \\
    \mathcal{F}_\gamma &= \frac{1}{4}\left( 1 + \langle Z_L^1Z_L^2 \rangle + \langle X_L^1Y_L^2 \rangle + \langle Y_L^1X_L^2 \rangle \right)
\end{align}
and are shown in Fig.~\ref{fig:FT_cnot}b as results of $N=10^6$ stabilizer simulation runs of the logical CNOT circuit followed by destructive measurement of all data qubits and ideal EC in the respective Pauli basis. Averaging over measurement results for the logical operators yields their expectation value.

The logical magic state $\ket{H}_L$ may be denoted by the logical density operator
\begin{align}
    \rho_H &= \ket{H}\bra{H}_L \simeq \frac{1}{2} \begin{pmatrix}
    1+ 1/\sqrt{2} & 1/\sqrt{2} \\
    1/\sqrt{2} & 1- 1/\sqrt{2}
    \end{pmatrix} \label{eq:rho_H}
\end{align}
where the matrix representation is in the logical computational basis. Quantum state tomography of the experimentally prepared logical magic state is shown in comparison to the theoretical values in Eq.~(\ref{eq:rho_H}) in Fig.~\ref{fig:FT_t_gen}b. 
The fidelity of the logical magic state as shown in Fig.~\ref{fig:FT_t_gen}c is given by
\begin{align}
    \mathcal{F}_H &= \frac{1}{2}\left( 1 + \frac{\langle X_L \rangle + \langle Z_L \rangle}{\sqrt{2}} \right)
\end{align}
since the logical magic state is the $+1$-eigenstate of the logical Hadamard operator $H_L\ket{H}_L = \ket{H}_L$ and its projector reads
\begin{align}
    P_H &= \frac{1}{2}\left(I + H_L\right) = \frac{1}{2}\left( I + \frac{X_L + Z_L }{\sqrt{2}} \right). \label{eq:projH}
\end{align}

When we inject the logical magic state onto logical Pauli states the result is the logical T-gate applied to the previously prepared logical Pauli state 
\begin{align}
    \ket{\psi}_{L,\text{out}} &= T_L \ket{t}_{L,\text{in}}\text{.}
\end{align}
For the four different input logical Pauli states $\ket{0}_L,\,\ket{1}_L,\,\ket{+}_L$ and $\ket{+i}_L$ the output states are
\begin{align}
    \ket{H}_L &= T_L \ket{0}_L \\ 
    \ket{-H}_L &= T_L \ket{1}_L \\ 
    X_L \ket{H}_L &= T_L \ket{+}_L \\ 
    \ket{+i}_L &= T_L \ket{+i}_L
\end{align}
and their projectors read
\begin{align}
    P_{0/1} &= \frac{1}{2}\left(I \pm H_L\right) = \frac{1}{2}\left( 1 \pm \frac{ X_L  +  Z_L }{\sqrt{2}} \right) \\
    P_+ &= X_L P_0 X_L = \frac{1}{2}\left( 1 + \frac{ X_L  -  Z_L }{\sqrt{2}} \right)  \\
    P_{+i} &= \frac{1}{2}\left(I + Y_L\right) \text{.}
\end{align}
The respective logical T-gate output state fidelities $\mathcal{F}_t$ for input Pauli state $\ket{t}_L$ are then given by
\begin{align}
    \mathcal{F}_0 &= \frac{1}{2}\left( 1 + \frac{\langle X_L \rangle + \langle Z_L \rangle}{\sqrt{2}} \right) \\
    \mathcal{F}_1 &= \frac{1}{2}\left( 1 - \frac{\langle X_L \rangle + \langle Z_L \rangle}{\sqrt{2}} \right) \\
    \mathcal{F}_+ &= \frac{1}{2}\left( 1 + \frac{\langle X_L \rangle - \langle Z_L \rangle}{\sqrt{2}} \right) \\
    \mathcal{F}_{+i} &= \frac{1}{2}\left( 1 + \langle Y_L \rangle \right)\text{.}
\end{align}

To estimate the expectation values of the logical operators occurring in the expressions for the fidelities given above, we run $N=10^5$ statevector simulations of the FT preparation and injection circuits. Each run is followed by destructive measurement in the respective Pauli basis and ideal EC that determines a measurement outcome of the logical Pauli operator. The expectation value is then calculated as the mean over all measurement outcomes.

The sampling uncertainty $\varepsilon_L$ when sampling the expectation value of a logical Pauli operator $O_L$ is \mbox{$\varepsilon_L = \sqrt{\frac{\text{Var}(\langle O_L \rangle)}{N}}$} and is propagated to their respective fidelities by Gaussian error propagation.

\subsection{Logical process matrix}
\label{sec:process_matrix}
Process matrices can be used to parameterize quantum channels
\begin{align}
    \mathcal{E}(\rho) &= \sum_{n=0}^{3}\sum_{m=0}^{3} \chi_{mn} E_m \rho E_n^\dagger
\end{align}
in the quantum operations formalism. The process matrix $\chi_{mn}$ for the logical T-gate that we show in Fig.~\ref{fig:FT_ti_inj}b is described by the quantum channel 
\begin{align}
    \mathcal{E}(\rho) &= T\rho T^\dagger = \sum_{n=0}^{3}\sum_{m=0}^{3} \chi_{mn} \sigma_m \rho \sigma_n
\end{align}
where we expand the channel in terms of the logical Pauli matrices. The matrix representation of $\chi$ in the logical Pauli basis reads
\begin{align}
    \chi &= \frac{1}{2}\begin{pmatrix}
    1+1/\sqrt{2} & 0 & i/\sqrt{2} & 0\\
    0 & 0 & 0 & 0\\
    -i/\sqrt{2} & 0 & 1-1/\sqrt{2} & 0\\
    0 & 0 & 0 & 0
    \end{pmatrix}  \text{.}
\end{align}
Measurements of expectation values of the logical Pauli basis for the \mbox{T-gate} input states $\ket{0}_L$, $\ket{1}_L$, $\ket{+}_L$ and $\ket{+i}_L$ form a tomographically complete set and allow for the reconstruction of the process matrix $\chi_{mn}$~\cite{Hradil2004}.

\subsection{Acceptance rates}
\label{sec:acceptance_rate}
We define the \emph{acceptance rate} as ratio of circuit runs where all flag qubits are measured as $+1$. The logical Pauli $\ket{0}_L$-state is fault-tolerantly encoded ($\ket{0}_{L,\text{ft}}$) using the circuit given in Fig.~\ref{fig:FT_prep}a, the logical magic state is prepared both by using the non-FT circuit followed by only the transversal Hadamard measurement ($\ket{H}_{L,\text{nf}}$ \& $M_H$) and by using the full FT protocol ($\ket{H}_{L,\text{ft}}$) as given in Fig.~\ref{fig:FT_t_gen}a. Approximate acceptance rates in simulation and experiment are shown in the table below for these three different encoding circuits alongside with the respective number of qubits acting as flags as well as circuit depth:

\begin{center}
\begin{tabular}{ c |c |c || c |c | c}
encoding & depth & \#flags & simulation & experiment & $ \Delta\varepsilon$\\ \hline 
$\ket{0}_{L,\text{ft}}$ & 25 & 1 & 85\% & 79\% & 7\%\\ 
$\ket{H}_{L,\text{nf}}$ \,\&\,$M_H$ & 32 & 2 & 72\% & 57\% & 21\%\\\  
$\ket{H}_{L,\text{ft}}$ & 75 & 8 & 27\% & 14\% & 48\%   
\end{tabular}
\end{center}

We observe that the relative error $\Delta \varepsilon$ between MC simulation and experimentally measured acceptance rates for the respective circuits increases with larger circuit depth and number of flag qubits involved.

\section{Data availability}
The data underlying the findings of this work and the quantum circuits are available at https://doi.org/10.5281/zenodo.5725062.

\section{Code availability}
All codes used for data analysis are available from the corresponding author upon reasonable request.

\section{Author contributions}
L.P., I.P. and T.F. carried out the experiments. L.P., I.P., T.F., M.Meth, C.D.M., R.S., M.Ringbauer, P.S. and T.M. contributed to the experimental setup. L.P. analyzed the data. S.H. performed the numerical simulations. S.H., M.Rispler and M.Müller performed circuit analysis, characterization and theory modelling. L.P., S.H., I.P., M.Rispler, P.S. and M.Müller wrote the manuscript, with contributions from all authors. R.B., P.S., M.Müller and T.M. supervised the project.

\section{Acknowledgements}
We acknowledge support from the EU Quantum Technology Flagship grant AQTION under grant agreement number 820495, and by the US Army Research Office through grant number W911NF-21-1-0007; and funding by the Austrian Science Fund (FWF), through the SFB BeyondC (FWF project number F7109), by the Austrian Research Promotion Agency (FFG) contract 872766, and by the IQI GmbH.
M. Ringbauer acknowledges funding from the European Union’s Horizon 2020 research and innovation programme under the Marie Skłodowska-Curie grant agreement number 840450.
M. Müller acknowledges support by the ERC Starting Grant QNets grant number 804247.
S.H. acknowledges funding by the Deutsche Forschungsgemeinschaft (DFG, German Research Foundation) under Germany’s Excellence Strategy ‘Cluster of Excellence Matter and Light for Quantum Computing (ML4Q) EXC 2004/1’ 390534769.
The research is also based on work supported by the Office of the Director of National Intelligence (ODNI), Intelligence Advanced Research Projects Activity (IARPA), via the US Army Research Office grant number W911NF-16-1-0070. The views and conclusions contained herein are those of the authors and should not be interpreted as necessarily representing the official policies or endorsements, either expressed or implied, of the ODNI, IARPA or the US Government. The US Government is authorized to reproduce and distribute reprints for governmental purposes notwithstanding any copyright annotation thereon. Any opinions, findings, and conclusions or recommendations expressed in this material are those of the author(s) and do not necessarily reflect the view of the US Army Research Office.

\bibliography{references}

\begin{thebibliography}{42}%
\makeatletter
\providecommand \@ifxundefined [1]{%
 \@ifx{#1\undefined}
}%
\providecommand \@ifnum [1]{%
 \ifnum #1\expandafter \@firstoftwo
 \else \expandafter \@secondoftwo
 \fi
}%
\providecommand \@ifx [1]{%
 \ifx #1\expandafter \@firstoftwo
 \else \expandafter \@secondoftwo
 \fi
}%
\providecommand \natexlab [1]{#1}%
\providecommand \enquote  [1]{``#1''}%
\providecommand \bibnamefont  [1]{#1}%
\providecommand \bibfnamefont [1]{#1}%
\providecommand \citenamefont [1]{#1}%
\providecommand \href@noop [0]{\@secondoftwo}%
\providecommand \href [0]{\begingroup \@sanitize@url \@href}%
\providecommand \@href[1]{\@@startlink{#1}\@@href}%
\providecommand \@@href[1]{\endgroup#1\@@endlink}%
\providecommand \@sanitize@url [0]{\catcode `\\12\catcode `\$12\catcode
  `\&12\catcode `\#12\catcode `\^12\catcode `\_12\catcode `\%12\relax}%
\providecommand \@@startlink[1]{}%
\providecommand \@@endlink[0]{}%
\providecommand \url  [0]{\begingroup\@sanitize@url \@url }%
\providecommand \@url [1]{\endgroup\@href {#1}{\urlprefix }}%
\providecommand \urlprefix  [0]{URL }%
\providecommand \Eprint [0]{\href }%
\providecommand \doibase [0]{https://doi.org/}%
\providecommand \selectlanguage [0]{\@gobble}%
\providecommand \bibinfo  [0]{\@secondoftwo}%
\providecommand \bibfield  [0]{\@secondoftwo}%
\providecommand \translation [1]{[#1]}%
\providecommand \BibitemOpen [0]{}%
\providecommand \bibitemStop [0]{}%
\providecommand \bibitemNoStop [0]{.\EOS\space}%
\providecommand \EOS [0]{\spacefactor3000\relax}%
\providecommand \BibitemShut  [1]{\csname bibitem#1\endcsname}%
\let\auto@bib@innerbib\@empty
\bibitem [{\citenamefont {Shor}(1997)}]{Shor1997}%
  \BibitemOpen
  \bibfield  {author} {\bibinfo {author} {P.~W. Shor},\ }\emph {Polynomial-Time
  Algorithms for Prime Factorization and Discrete Logarithms on a Quantum
  Computer},\ \href {https://doi.org/10.1137/S0097539795293172} {\bibfield
  {journal} {\bibinfo  {journal} {SIAM Journal on Computing}\ }\textbf
  {\bibinfo {volume} {26}},\ \bibinfo {pages} {1484} (\bibinfo {year}
  {1997})}\BibitemShut {NoStop}%
\bibitem [{\citenamefont {Feynman}(1982)}]{Feynman1982}%
  \BibitemOpen
  \bibfield  {author} {\bibinfo {author} {R.~P. Feynman},\ }\emph {{Simulating
  physics with computers}},\ \href {https://doi.org/10.1007/BF02650179}
  {\bibfield  {journal} {\bibinfo  {journal} {Int. J. Theor. Phys.}\ }\textbf
  {\bibinfo {volume} {21}},\ \bibinfo {pages} {467} (\bibinfo {year}
  {1982})}\BibitemShut {NoStop}%
\bibitem [{\citenamefont {Nielsen}\ and\ \citenamefont
  {Chuang}(2010)}]{Nielsen2010}%
  \BibitemOpen
  \bibfield  {author} {\bibinfo {author} {M.~A. Nielsen}\ and\ \bibinfo
  {author} {I.~L. Chuang},\ }\href {https://doi.org/10.1017/CBO9780511976667}
  {\emph {\bibinfo {title} {Quantum Computation and Quantum Information: 10th
  Anniversary Edition}}}\ (\bibinfo  {publisher} {Cambridge University Press},\
  \bibinfo {year} {2010})\BibitemShut {NoStop}%
\bibitem [{\citenamefont {Shor}(1996)}]{Shor1996}%
  \BibitemOpen
  \bibfield  {author} {\bibinfo {author} {P.~W. Shor},\ }in\ \href
  {https://doi.org/10.1109/SFCS.1996.548464} {\emph {\bibinfo {booktitle}
  {Proceedings of 37th Conference on Foundations of Computer Science}}}\
  (\bibinfo  {publisher} {IEEE},\ \bibinfo {year} {1996})\ pp.\ \bibinfo
  {pages} {56--65}\BibitemShut {NoStop}%
\bibitem [{\citenamefont {Preskill}(1998)}]{Preskill1998}%
  \BibitemOpen
  \bibfield  {author} {\bibinfo {author} {J.~Preskill},\ }\emph {{Reliable
  quantum computers}},\ \href {https://doi.org/10.1098/rspa.1998.0167}
  {\bibfield  {journal} {\bibinfo  {journal} {Proc. R. Soc. Lond. A.}\ }\textbf
  {\bibinfo {volume} {454}},\ \bibinfo {pages} {385} (\bibinfo {year}
  {1998})}\BibitemShut {NoStop}%
\bibitem [{\citenamefont {Aliferis}\ \emph {et~al.}(2006)\citenamefont
  {Aliferis}, \citenamefont {Gottesman},\ and\ \citenamefont
  {Preskill}}]{Aliferis2006}%
  \BibitemOpen
  \bibfield  {author} {\bibinfo {author} {P.~Aliferis}, \bibinfo {author}
  {D.~Gottesman},\ and\ \bibinfo {author} {J.~Preskill},\ }\emph {{Quantum
  accuracy threshold for concatenated distance-3 codes}},\ \href@noop {}
  {\bibfield  {journal} {\bibinfo  {journal} {Quantum Inf. Comput.}\ }\textbf
  {\bibinfo {volume} {6}},\ \bibinfo {pages} {97} (\bibinfo {year}
  {2006})}\BibitemShut {NoStop}%
\bibitem [{\citenamefont {Terhal}(2015)}]{Terhal2015}%
  \BibitemOpen
  \bibfield  {author} {\bibinfo {author} {B.~M. Terhal},\ }\emph {{Quantum
  error correction for quantum memories}},\ \href
  {https://doi.org/10.1103/RevModPhys.87.307} {\bibfield  {journal} {\bibinfo
  {journal} {Rev. Mod. Phys.}\ }\textbf {\bibinfo {volume} {87}},\ \bibinfo
  {pages} {307} (\bibinfo {year} {2015})}\BibitemShut {NoStop}%
\bibitem [{\citenamefont {Aharonov}\ and\ \citenamefont
  {Ben-Or}(2008)}]{Aharonov2008}%
  \BibitemOpen
  \bibfield  {author} {\bibinfo {author} {D.~Aharonov}\ and\ \bibinfo {author}
  {M.~Ben-Or},\ }\emph {Fault-Tolerant Quantum Computation with Constant Error
  Rate},\ \href {https://doi.org/10.1137/S0097539799359385} {\bibfield
  {journal} {\bibinfo  {journal} {SIAM Journal on Computing}\ }\textbf
  {\bibinfo {volume} {38}},\ \bibinfo {pages} {1207} (\bibinfo {year}
  {2008})}\BibitemShut {NoStop}%
\bibitem [{\citenamefont {Eastin}\ and\ \citenamefont
  {Knill}(2009)}]{Eastin2009}%
  \BibitemOpen
  \bibfield  {author} {\bibinfo {author} {B.~Eastin}\ and\ \bibinfo {author}
  {E.~Knill},\ }\emph {{Restrictions on Transversal Encoded Quantum Gate
  Sets}},\ \href {https://doi.org/10.1103/PhysRevLett.102.110502} {\bibfield
  {journal} {\bibinfo  {journal} {Phys. Rev. Lett.}\ }\textbf {\bibinfo
  {volume} {102}},\ \bibinfo {pages} {110502} (\bibinfo {year}
  {2009})}\BibitemShut {NoStop}%
\bibitem [{\citenamefont {Bravyi}\ and\ \citenamefont
  {Kitaev}(2005)}]{Bravyi2005}%
  \BibitemOpen
  \bibfield  {author} {\bibinfo {author} {S.~Bravyi}\ and\ \bibinfo {author}
  {A.~Kitaev},\ }\emph {{Universal quantum computation with ideal Clifford
  gates and noisy ancillas}},\ \href
  {https://doi.org/10.1103/PhysRevA.71.022316} {\bibfield  {journal} {\bibinfo
  {journal} {Phys. Rev. A}\ }\textbf {\bibinfo {volume} {71}},\ \bibinfo
  {pages} {022316} (\bibinfo {year} {2005})}\BibitemShut {NoStop}%
\bibitem [{\citenamefont {Paetznick}\ and\ \citenamefont
  {Reichardt}(2013)}]{Paetznick2013}%
  \BibitemOpen
  \bibfield  {author} {\bibinfo {author} {A.~Paetznick}\ and\ \bibinfo {author}
  {B.~W. Reichardt},\ }\emph {{Universal Fault-Tolerant Quantum Computation
  with Only Transversal Gates and Error Correction}},\ \href
  {https://doi.org/10.1103/PhysRevLett.111.090505} {\bibfield  {journal}
  {\bibinfo  {journal} {Phys. Rev. Lett.}\ }\textbf {\bibinfo {volume} {111}},\
  \bibinfo {pages} {090505} (\bibinfo {year} {2013})}\BibitemShut {NoStop}%
\bibitem [{\citenamefont {Beverland}\ \emph {et~al.}(2021)\citenamefont
  {Beverland}, \citenamefont {Kubica},\ and\ \citenamefont
  {Svore}}]{Beverland2021}%
  \BibitemOpen
  \bibfield  {author} {\bibinfo {author} {M.~E. Beverland}, \bibinfo {author}
  {A.~Kubica},\ and\ \bibinfo {author} {K.~M. Svore},\ }\emph {{Cost of
  Universality: A Comparative Study of the Overhead of State Distillation and
  Code Switching with Color Codes}},\ \href
  {https://doi.org/10.1103/PRXQuantum.2.020341} {\bibfield  {journal} {\bibinfo
   {journal} {PRX Quantum}\ }\textbf {\bibinfo {volume} {2}},\ \bibinfo {pages}
  {020341} (\bibinfo {year} {2021})}\BibitemShut {NoStop}%
\bibitem [{\citenamefont {Nigg}\ \emph {et~al.}(2014)\citenamefont {Nigg},
  \citenamefont {M{\ifmmode\ddot{u}\else\"{u}\fi}ller}, \citenamefont
  {Martinez}, \citenamefont {Schindler}, \citenamefont {Hennrich},
  \citenamefont {Monz}, \citenamefont {Martin-Delgado},\ and\ \citenamefont
  {Blatt}}]{Nigg2014}%
  \BibitemOpen
  \bibfield  {author} {\bibinfo {author} {D.~Nigg}, et~al.,\ }\emph {{Quantum
  computations on a topologically encoded qubit}},\ \href
  {https://doi.org/10.1126/science.1253742} {\bibfield  {journal} {\bibinfo
  {journal} {Science}\ }\textbf {\bibinfo {volume} {345}},\ \bibinfo {pages}
  {302} (\bibinfo {year} {2014})}\BibitemShut {NoStop}%
\bibitem [{\citenamefont {Satzinger}\ \emph {et~al.}(2021)\citenamefont
  {Satzinger}, \citenamefont {Liu}, \citenamefont {Smith}, \citenamefont
  {Knapp}, \citenamefont {Newman}, \citenamefont {Jones}, \citenamefont {Chen},
  \citenamefont {Quintana}, \citenamefont {Mi}, \citenamefont {Dunsworth},
  \citenamefont {Gidney}, \citenamefont {Aleiner}, \citenamefont {Arute},
  \citenamefont {Arya}, \citenamefont {Atalaya}, \citenamefont {Babbush},
  \citenamefont {Bardin}, \citenamefont {Barends}, \citenamefont {Basso},
  \citenamefont {Bengtsson}, \citenamefont {Bilmes}, \citenamefont {Broughton},
  \citenamefont {Buckley}, \citenamefont {Buell}, \citenamefont {Burkett},
  \citenamefont {Bushnell}, \citenamefont {Chiaro}, \citenamefont {Collins},
  \citenamefont {Courtney}, \citenamefont {Demura}, \citenamefont {Derk},
  \citenamefont {Eppens}, \citenamefont {Erickson}, \citenamefont {Farhi},
  \citenamefont {Foaro}, \citenamefont {Fowler}, \citenamefont {Foxen},
  \citenamefont {Giustina}, \citenamefont {Greene}, \citenamefont {Gross},
  \citenamefont {Harrigan}, \citenamefont {Harrington}, \citenamefont {Hilton},
  \citenamefont {Hong}, \citenamefont {Huang}, \citenamefont {Huggins},
  \citenamefont {Ioffe}, \citenamefont {Isakov}, \citenamefont {Jeffrey},
  \citenamefont {Jiang}, \citenamefont {Kafri}, \citenamefont {Kechedzhi},
  \citenamefont {Khattar}, \citenamefont {Kim}, \citenamefont {Klimov},
  \citenamefont {Korotkov}, \citenamefont {Kostritsa}, \citenamefont
  {Landhuis}, \citenamefont {Laptev}, \citenamefont {Locharla}, \citenamefont
  {Lucero}, \citenamefont {Martin}, \citenamefont {McClean}, \citenamefont
  {McEwen}, \citenamefont {Miao}, \citenamefont {Mohseni}, \citenamefont
  {Montazeri}, \citenamefont {Mruczkiewicz}, \citenamefont {Mutus},
  \citenamefont {Naaman}, \citenamefont {Neeley}, \citenamefont {Neill},
  \citenamefont {Niu}, \citenamefont {O'Brien}, \citenamefont {Opremcak},
  \citenamefont {Pat{\ifmmode\acute{o}\else\'{o}\fi}}, \citenamefont
  {Petukhov}, \citenamefont {Rubin}, \citenamefont {Sank}, \citenamefont
  {Shvarts}, \citenamefont {Strain}, \citenamefont {Szalay}, \citenamefont
  {Villalonga}, \citenamefont {White}, \citenamefont {Yao}, \citenamefont
  {Yeh}, \citenamefont {Yoo}, \citenamefont {Zalcman}, \citenamefont {Neven},
  \citenamefont {Boixo}, \citenamefont {Megrant}, \citenamefont {Chen},
  \citenamefont {Kelly}, \citenamefont {Smelyanskiy}, \citenamefont {Kitaev},
  \citenamefont {Knap}, \citenamefont {Pollmann},\ and\ \citenamefont
  {Roushan}}]{Satzinger2021}%
  \BibitemOpen
  \bibfield  {author} {\bibinfo {author} {K.~J. Satzinger}, et~al.,\ }\emph
  {{Realizing topologically ordered states on a quantum processor}},\ \href
  {https://arxiv.org/abs/2104.01180v1} {\bibfield  {journal} {\bibinfo
  {journal} {arXiv}\ } (\bibinfo {year} {2021})},\ \Eprint
  {https://arxiv.org/abs/2104.01180} {2104.01180} \BibitemShut {NoStop}%
\bibitem [{\citenamefont {Andersen}\ \emph {et~al.}(2020)\citenamefont
  {Andersen}, \citenamefont {Remm}, \citenamefont {Lazar}, \citenamefont
  {Krinner}, \citenamefont {Lacroix}, \citenamefont {Norris}, \citenamefont
  {Gabureac}, \citenamefont {Eichler},\ and\ \citenamefont
  {Wallraff}}]{Andersen2020}%
  \BibitemOpen
  \bibfield  {author} {\bibinfo {author} {C.~K. Andersen}, et~al.,\ }\emph
  {Repeated quantum error detection in a surface code},\ \href
  {https://doi.org/10.1038/s41567-020-0920-y} {\bibfield  {journal} {\bibinfo
  {journal} {Nature Physics}\ }\textbf {\bibinfo {volume} {16}},\ \bibinfo
  {pages} {875} (\bibinfo {year} {2020})}\BibitemShut {NoStop}%
\bibitem [{\citenamefont {Marques}\ \emph {et~al.}(2021)\citenamefont
  {Marques}, \citenamefont {Varbanov}, \citenamefont {Moreira}, \citenamefont
  {Ali}, \citenamefont {Muthusubramanian}, \citenamefont {Zachariadis},
  \citenamefont {Battistel}, \citenamefont {Beekman}, \citenamefont {Haider},
  \citenamefont {Vlothuizen}, \citenamefont {Bruno}, \citenamefont {Terhal},\
  and\ \citenamefont {DiCarlo}}]{Marques2021}%
  \BibitemOpen
  \bibfield  {author} {\bibinfo {author} {J.~F. Marques}, et~al.,\ }\emph
  {{Logical-qubit operations in an error-detecting surface code}},\ \href
  {https://arxiv.org/abs/2102.13071v1} {\bibfield  {journal} {\bibinfo
  {journal} {arXiv}\ } (\bibinfo {year} {2021})},\ \Eprint
  {https://arxiv.org/abs/2102.13071} {2102.13071} \BibitemShut {NoStop}%
\bibitem [{\citenamefont {Chen}\ \emph {et~al.}(2021)\citenamefont {Chen},
  \citenamefont {Satzinger}, \citenamefont {Atalaya}, \citenamefont {Korotkov},
  \citenamefont {Dunsworth}, \citenamefont {Sank}, \citenamefont {Quintana},
  \citenamefont {McEwen}, \citenamefont {Barends}, \citenamefont {Klimov},
  \citenamefont {Hong}, \citenamefont {Jones}, \citenamefont {Petukhov},
  \citenamefont {Kafri}, \citenamefont {Demura}, \citenamefont {Burkett},
  \citenamefont {Gidney}, \citenamefont {Fowler}, \citenamefont {Paler},
  \citenamefont {Putterman}, \citenamefont {Aleiner}, \citenamefont {Arute},
  \citenamefont {Arya}, \citenamefont {Babbush}, \citenamefont {Bardin},
  \citenamefont {Bengtsson}, \citenamefont {Bourassa}, \citenamefont
  {Broughton}, \citenamefont {Buckley}, \citenamefont {Buell}, \citenamefont
  {Bushnell}, \citenamefont {Chiaro}, \citenamefont {Collins}, \citenamefont
  {Courtney}, \citenamefont {Derk}, \citenamefont {Eppens}, \citenamefont
  {Erickson}, \citenamefont {Farhi}, \citenamefont {Foxen}, \citenamefont
  {Giustina}, \citenamefont {Greene}, \citenamefont {Gross}, \citenamefont
  {Harrigan}, \citenamefont {Harrington}, \citenamefont {Hilton}, \citenamefont
  {Ho}, \citenamefont {Huang}, \citenamefont {Huggins}, \citenamefont {Ioffe},
  \citenamefont {Isakov}, \citenamefont {Jeffrey}, \citenamefont {Jiang},
  \citenamefont {Kechedzhi}, \citenamefont {Kim}, \citenamefont {Kitaev},
  \citenamefont {Kostritsa}, \citenamefont {Landhuis}, \citenamefont {Laptev},
  \citenamefont {Lucero}, \citenamefont {Martin}, \citenamefont {McClean},
  \citenamefont {McCourt}, \citenamefont {Mi}, \citenamefont {Miao},
  \citenamefont {Mohseni}, \citenamefont {Montazeri}, \citenamefont
  {Mruczkiewicz}, \citenamefont {Mutus}, \citenamefont {Naaman}, \citenamefont
  {Neeley}, \citenamefont {Neill}, \citenamefont {Newman}, \citenamefont {Niu},
  \citenamefont {O{'}Brien}, \citenamefont {Opremcak}, \citenamefont {Ostby},
  \citenamefont {Pat{\ifmmode\acute{o}\else\'{o}\fi}}, \citenamefont {Redd},
  \citenamefont {Roushan}, \citenamefont {Rubin}, \citenamefont {Shvarts},
  \citenamefont {Strain}, \citenamefont {Szalay}, \citenamefont {Trevithick},
  \citenamefont {Villalonga}, \citenamefont {White}, \citenamefont {Yao},
  \citenamefont {Yeh}, \citenamefont {Yoo}, \citenamefont {Zalcman},
  \citenamefont {Neven}, \citenamefont {Boixo}, \citenamefont {Smelyanskiy},
  \citenamefont {Chen}, \citenamefont {Megrant}, \citenamefont {Kelly},\ and\
  \citenamefont {A.~I.}}]{Chen2021}%
  \BibitemOpen
  \bibfield  {author} {\bibinfo {author} {Z.~Chen}, et~al.,\ }\emph
  {{Exponential suppression of bit or phase errors with cyclic error
  correction}},\ \href {https://doi.org/10.1038/s41586-021-03588-y} {\bibfield
  {journal} {\bibinfo  {journal} {Nature}\ }\textbf {\bibinfo {volume} {595}},\
  \bibinfo {pages} {383} (\bibinfo {year} {2021})}\BibitemShut {NoStop}%
\bibitem [{\citenamefont {Gottesman}(2016)}]{Gottesman2016}%
  \BibitemOpen
  \bibfield  {author} {\bibinfo {author} {D.~Gottesman},\ }\emph {{Quantum
  fault tolerance in small experiments}},\ \href
  {https://arxiv.org/abs/1610.03507v2} {\bibfield  {journal} {\bibinfo
  {journal} {arXiv}\ } (\bibinfo {year} {2016})},\ \Eprint
  {https://arxiv.org/abs/1610.03507} {1610.03507} \BibitemShut {NoStop}%
\bibitem [{\citenamefont {Takita}\ \emph {et~al.}(2017)\citenamefont {Takita},
  \citenamefont {Cross}, \citenamefont
  {C{\ifmmode\acute{o}\else\'{o}\fi}rcoles}, \citenamefont {Chow},\ and\
  \citenamefont {Gambetta}}]{Takita2017}%
  \BibitemOpen
  \bibfield  {author} {\bibinfo {author} {M.~Takita}, \bibinfo {author} {A.~W.
  Cross}, \bibinfo {author} {A.~D. C{\ifmmode\acute{o}\else\'{o}\fi}rcoles},
  \bibinfo {author} {J.~M. Chow},\ and\ \bibinfo {author} {J.~M. Gambetta},\
  }\emph {{Experimental Demonstration of Fault-Tolerant State Preparation with
  Superconducting Qubits}},\ \href
  {https://doi.org/10.1103/PhysRevLett.119.180501} {\bibfield  {journal}
  {\bibinfo  {journal} {Phys. Rev. Lett.}\ }\textbf {\bibinfo {volume} {119}},\
  \bibinfo {pages} {180501} (\bibinfo {year} {2017})}\BibitemShut {NoStop}%
\bibitem [{\citenamefont {Vuillot}(2018)}]{Vuillot2018}%
  \BibitemOpen
  \bibfield  {author} {\bibinfo {author} {C.~Vuillot},\ }\emph {{Is error
  detection helpful on IBM 5Q chips?}},\ \href@noop {} {\bibfield  {journal}
  {\bibinfo  {journal} {Quantum Inf. Comput.}\ }\textbf {\bibinfo {volume}
  {18}},\ \bibinfo {pages} {949} (\bibinfo {year} {2018})}\BibitemShut
  {NoStop}%
\bibitem [{\citenamefont {Linke}\ \emph {et~al.}(2017)\citenamefont {Linke},
  \citenamefont {Gutierrez}, \citenamefont {Landsman}, \citenamefont {Figgatt},
  \citenamefont {Debnath}, \citenamefont {Brown},\ and\ \citenamefont
  {Monroe}}]{Linke2017}%
  \BibitemOpen
  \bibfield  {author} {\bibinfo {author} {N.~M. Linke}, et~al.,\ }\emph
  {Fault-tolerant quantum error detection},\ \href
  {https://doi.org/10.1126/sciadv.1701074} {\bibfield  {journal} {\bibinfo
  {journal} {Sci. Adv.}\ }\textbf {\bibinfo {volume} {3}},\ \bibinfo {pages}
  {e1701074} (\bibinfo {year} {2017})}\BibitemShut {NoStop}%
\bibitem [{\citenamefont {Egan}\ \emph {et~al.}(2021)\citenamefont {Egan},
  \citenamefont {Debroy}, \citenamefont {Noel}, \citenamefont {Zhu},
  \citenamefont {Newman}, \citenamefont {Brown}, \citenamefont {Cetina},\ and\
  \citenamefont {Monroe}}]{Egan2021}%
  \BibitemOpen
  \bibfield  {author} {\bibinfo {author} {L.~Egan}, et~al.,\ }\emph
  {Fault-tolerant control of an error-corrected qubit},\ \href
  {https://doi.org/10.1038/s41586-021-03928-y} {\bibfield  {journal} {\bibinfo
  {journal} {Nature}\ }\textbf {\bibinfo {volume} {598}},\ \bibinfo {pages}
  {281} (\bibinfo {year} {2021})}\BibitemShut {NoStop}%
\bibitem [{\citenamefont {Chao}\ and\ \citenamefont
  {Reichardt}(2018)}]{Chao2018}%
  \BibitemOpen
  \bibfield  {author} {\bibinfo {author} {R.~Chao}\ and\ \bibinfo {author}
  {B.~W. Reichardt},\ }\emph {{Quantum Error Correction with Only Two Extra
  Qubits}},\ \href {https://doi.org/10.1103/PhysRevLett.121.050502} {\bibfield
  {journal} {\bibinfo  {journal} {Phys. Rev. Lett.}\ }\textbf {\bibinfo
  {volume} {121}},\ \bibinfo {pages} {050502} (\bibinfo {year}
  {2018})}\BibitemShut {NoStop}%
\bibitem [{\citenamefont {Chamberland}\ and\ \citenamefont
  {Beverland}(2018)}]{Chamberland2018}%
  \BibitemOpen
  \bibfield  {author} {\bibinfo {author} {C.~Chamberland}\ and\ \bibinfo
  {author} {M.~E. Beverland},\ }\emph {{Flag fault-tolerant error correction
  with arbitrary distance codes}},\ \href
  {https://doi.org/10.22331/q-2018-02-08-53} {\bibfield  {journal} {\bibinfo
  {journal} {Quantum}\ }\textbf {\bibinfo {volume} {2}},\ \bibinfo {pages} {53}
  (\bibinfo {year} {2018})}\BibitemShut {NoStop}%
\bibitem [{\citenamefont {Chamberland}\ and\ \citenamefont
  {Cross}(2019)}]{Chamberland2019}%
  \BibitemOpen
  \bibfield  {author} {\bibinfo {author} {C.~Chamberland}\ and\ \bibinfo
  {author} {A.~W. Cross},\ }\emph {{Fault-tolerant magic state preparation with
  flag qubits}},\ \href {https://doi.org/10.22331/q-2019-05-20-143} {\bibfield
  {journal} {\bibinfo  {journal} {Quantum}\ }\textbf {\bibinfo {volume} {3}},\
  \bibinfo {pages} {143} (\bibinfo {year} {2019})}\BibitemShut {NoStop}%
\bibitem [{\citenamefont {Chao}\ and\ \citenamefont
  {Reichardt}(2020)}]{Chao2020}%
  \BibitemOpen
  \bibfield  {author} {\bibinfo {author} {R.~Chao}\ and\ \bibinfo {author}
  {B.~W. Reichardt},\ }\emph {{Flag Fault-Tolerant Error Correction for any
  Stabilizer Code}},\ \href {https://doi.org/10.1103/PRXQuantum.1.010302}
  {\bibfield  {journal} {\bibinfo  {journal} {PRX Quantum}\ }\textbf {\bibinfo
  {volume} {1}},\ \bibinfo {pages} {010302} (\bibinfo {year}
  {2020})}\BibitemShut {NoStop}%
\bibitem [{\citenamefont {Reichardt}(2020)}]{Reichardt2020}%
  \BibitemOpen
  \bibfield  {author} {\bibinfo {author} {B.~W. Reichardt},\ }\emph
  {{Fault-tolerant quantum error correction for Steane{'}s seven-qubit color
  code with few or no extra qubits}},\ \href
  {https://doi.org/10.1088/2058-9565/abc6f4} {\bibfield  {journal} {\bibinfo
  {journal} {Quantum Sci. Technol.}\ }\textbf {\bibinfo {volume} {6}},\
  \bibinfo {pages} {015007} (\bibinfo {year} {2020})}\BibitemShut {NoStop}%
\bibitem [{\citenamefont {Abobeih}\ \emph {et~al.}(2021)\citenamefont
  {Abobeih}, \citenamefont {Wang}, \citenamefont {Randall}, \citenamefont
  {Loenen}, \citenamefont {Bradley}, \citenamefont {Markham}, \citenamefont
  {Twitchen}, \citenamefont {Terhal},\ and\ \citenamefont
  {Taminiau}}]{Abobeih2021}%
  \BibitemOpen
  \bibfield  {author} {\bibinfo {author} {M.~H. Abobeih}, et~al.,\ }\emph
  {{Fault-tolerant operation of a logical qubit in a diamond quantum
  processor}},\ \href {https://arxiv.org/abs/2108.01646v1} {\bibfield
  {journal} {\bibinfo  {journal} {arXiv}\ } (\bibinfo {year} {2021})},\ \Eprint
  {https://arxiv.org/abs/2108.01646} {2108.01646} \BibitemShut {NoStop}%
\bibitem [{\citenamefont {Hilder}\ \emph {et~al.}(2021)\citenamefont {Hilder},
  \citenamefont {Pijn}, \citenamefont {Onishchenko}, \citenamefont {Stahl},
  \citenamefont {Orth}, \citenamefont {Lekitsch}, \citenamefont
  {Rodriguez-Blanco}, \citenamefont {M{\ifmmode\ddot{u}\else\"{u}\fi}ller},
  \citenamefont {Schmidt-Kaler},\ and\ \citenamefont
  {Poschinger}}]{Hilder2021}%
  \BibitemOpen
  \bibfield  {author} {\bibinfo {author} {J.~Hilder}, et~al.,\ }\emph
  {{Fault-tolerant parity readout on a shuttling-based trapped-ion quantum
  computer}},\ \href {https://arxiv.org/abs/2107.06368v1} {\bibfield  {journal}
  {\bibinfo  {journal} {arXiv}\ } (\bibinfo {year} {2021})},\ \Eprint
  {https://arxiv.org/abs/2107.06368} {2107.06368} \BibitemShut {NoStop}%
\bibitem [{\citenamefont {Ryan-Anderson}\ \emph {et~al.}(2021)\citenamefont
  {Ryan-Anderson}, \citenamefont {Bohnet}, \citenamefont {Lee}, \citenamefont
  {Gresh}, \citenamefont {Hankin}, \citenamefont {Gaebler}, \citenamefont
  {Francois}, \citenamefont {Chernoguzov}, \citenamefont {Lucchetti},
  \citenamefont {Brown}, \citenamefont {Gatterman}, \citenamefont {Halit},
  \citenamefont {Gilmore}, \citenamefont {Gerber}, \citenamefont {Neyenhuis},
  \citenamefont {Hayes},\ and\ \citenamefont {Stutz}}]{Ryan-Anderson2021}%
  \BibitemOpen
  \bibfield  {author} {\bibinfo {author} {C.~Ryan-Anderson}, et~al.,\ }\emph
  {{Realization of real-time fault-tolerant quantum error correction}},\ \href
  {https://arxiv.org/abs/2107.07505v1} {\bibfield  {journal} {\bibinfo
  {journal} {arXiv}\ } (\bibinfo {year} {2021})},\ \Eprint
  {https://arxiv.org/abs/2107.07505} {2107.07505} \BibitemShut {NoStop}%
\bibitem [{\citenamefont {Bombin}\ and\ \citenamefont
  {Martin-Delgado}(2006)}]{Bombin2006}%
  \BibitemOpen
  \bibfield  {author} {\bibinfo {author} {H.~Bombin}\ and\ \bibinfo {author}
  {M.~A. Martin-Delgado},\ }\emph {{Topological Quantum Distillation}},\ \href
  {https://doi.org/10.1103/PhysRevLett.97.180501} {\bibfield  {journal}
  {\bibinfo  {journal} {Phys. Rev. Lett.}\ }\textbf {\bibinfo {volume} {97}},\
  \bibinfo {pages} {180501} (\bibinfo {year} {2006})}\BibitemShut {NoStop}%
\bibitem [{\citenamefont {Steane}(1996)}]{Steane1996}%
  \BibitemOpen
  \bibfield  {author} {\bibinfo {author} {A.~Steane},\ }\emph
  {{Multiple-particle interference and quantum error correction}},\ \href
  {https://doi.org/10.1098/rspa.1996.0136} {\bibfield  {journal} {\bibinfo
  {journal} {Proc. R. Soc. Lond. A.}\ }\textbf {\bibinfo {volume} {452}},\
  \bibinfo {pages} {2551} (\bibinfo {year} {1996})}\BibitemShut {NoStop}%
\bibitem [{\citenamefont {Pogorelov}\ \emph {et~al.}(2021)\citenamefont
  {Pogorelov}, \citenamefont {Feldker}, \citenamefont {Marciniak},
  \citenamefont {Postler}, \citenamefont {Jacob}, \citenamefont
  {Krieglsteiner}, \citenamefont {Podlesnic}, \citenamefont {Meth},
  \citenamefont {Negnevitsky}, \citenamefont {Stadler}, \citenamefont
  {H\"ofer}, \citenamefont {W\"achter}, \citenamefont {Lakhmanskiy},
  \citenamefont {Blatt}, \citenamefont {Schindler},\ and\ \citenamefont
  {Monz}}]{Pogorelov2021}%
  \BibitemOpen
  \bibfield  {author} {\bibinfo {author} {I.~Pogorelov}, et~al.,\ }\emph
  {Compact Ion-Trap Quantum Computing Demonstrator},\ \href
  {https://doi.org/10.1103/PRXQuantum.2.020343} {\bibfield  {journal} {\bibinfo
   {journal} {PRX Quantum}\ }\textbf {\bibinfo {volume} {2}},\ \bibinfo {pages}
  {020343} (\bibinfo {year} {2021})}\BibitemShut {NoStop}%
\bibitem [{\citenamefont {S\o{}rensen}\ and\ \citenamefont
  {M\o{}lmer}(2000)}]{Sorensen2000}%
  \BibitemOpen
  \bibfield  {author} {\bibinfo {author} {A.~S\o{}rensen}\ and\ \bibinfo
  {author} {K.~M\o{}lmer},\ }\emph {Entanglement and quantum computation with
  ions in thermal motion},\ \href {https://doi.org/10.1103/PhysRevA.62.022311}
  {\bibfield  {journal} {\bibinfo  {journal} {Phys. Rev. A}\ }\textbf {\bibinfo
  {volume} {62}},\ \bibinfo {pages} {022311} (\bibinfo {year}
  {2000})}\BibitemShut {NoStop}%
\bibitem [{\citenamefont {Nebendahl}\ \emph {et~al.}(2009)\citenamefont
  {Nebendahl}, \citenamefont {H\"affner},\ and\ \citenamefont
  {Roos}}]{Nebendahl2009}%
  \BibitemOpen
  \bibfield  {author} {\bibinfo {author} {V.~Nebendahl}, \bibinfo {author}
  {H.~H\"affner},\ and\ \bibinfo {author} {C.~F. Roos},\ }\emph {Optimal
  control of entangling operations for trapped-ion quantum computing},\ \href
  {https://doi.org/10.1103/PhysRevA.79.012312} {\bibfield  {journal} {\bibinfo
  {journal} {Phys. Rev. A}\ }\textbf {\bibinfo {volume} {79}},\ \bibinfo
  {pages} {012312} (\bibinfo {year} {2009})}\BibitemShut {NoStop}%
\bibitem [{\citenamefont {Goto}(2016)}]{Goto2016}%
  \BibitemOpen
  \bibfield  {author} {\bibinfo {author} {H.~Goto},\ }\emph {Minimizing
  resource overheads for fault-tolerant preparation of encoded states of the
  Steane code},\ \href {https://doi.org/10.1038/srep19578} {\bibfield
  {journal} {\bibinfo  {journal} {Scientific Reports}\ }\textbf {\bibinfo
  {volume} {6}},\ \bibinfo {pages} {19578} (\bibinfo {year}
  {2016})}\BibitemShut {NoStop}%
\bibitem [{\citenamefont {Bermudez}\ \emph {et~al.}(2019)\citenamefont
  {Bermudez}, \citenamefont {Xu}, \citenamefont {Guti\'errez}, \citenamefont
  {Benjamin},\ and\ \citenamefont {M\"uller}}]{Bermudez2019}%
  \BibitemOpen
  \bibfield  {author} {\bibinfo {author} {A.~Bermudez}, \bibinfo {author}
  {X.~Xu}, \bibinfo {author} {M.~Guti\'errez}, \bibinfo {author} {S.~C.
  Benjamin},\ and\ \bibinfo {author} {M.~M\"uller},\ }\emph {Fault-tolerant
  protection of near-term trapped-ion topological qubits under realistic noise
  sources},\ \href {https://doi.org/10.1103/PhysRevA.100.062307} {\bibfield
  {journal} {\bibinfo  {journal} {Phys. Rev. A}\ }\textbf {\bibinfo {volume}
  {100}},\ \bibinfo {pages} {062307} (\bibinfo {year} {2019})}\BibitemShut
  {NoStop}%
\bibitem [{\citenamefont {Riesebos}\ \emph {et~al.}(2017)\citenamefont
  {Riesebos}, \citenamefont {Fu}, \citenamefont {Varsamopoulos}, \citenamefont
  {Almudever},\ and\ \citenamefont {Bertels}}]{riesebos2017pauli}%
  \BibitemOpen
  \bibfield  {author} {\bibinfo {author} {L.~Riesebos}, \bibinfo {author}
  {X.~Fu}, \bibinfo {author} {S.~Varsamopoulos}, \bibinfo {author} {C.~G.
  Almudever},\ and\ \bibinfo {author} {K.~Bertels},\ }in\ \href
  {https://doi.org/10.1145/3061639.3062300} {\emph {\bibinfo {booktitle} {{DAC
  '17: Proceedings of the 54th Annual Design Automation Conference 2017}}}}\
  (\bibinfo  {publisher} {Association for Computing Machinery},\ \bibinfo
  {address} {New York, NY, USA},\ \bibinfo {year} {2017})\ pp.\ \bibinfo
  {pages} {1--6}\BibitemShut {NoStop}%
\bibitem [{\citenamefont {Knill}(2005)}]{Knill2005}%
  \BibitemOpen
  \bibfield  {author} {\bibinfo {author} {E.~Knill},\ }\emph {{Quantum
  computing with realistically noisy devices - Nature}},\ \href
  {https://doi.org/10.1038/nature03350} {\bibfield  {journal} {\bibinfo
  {journal} {Nature}\ }\textbf {\bibinfo {volume} {434}},\ \bibinfo {pages}
  {39} (\bibinfo {year} {2005})}\BibitemShut {NoStop}%
\bibitem [{\citenamefont {Hradil}\ \emph {et~al.}(2004)\citenamefont {Hradil},
  \citenamefont {{\v{R}}eh{\'a}{\v{c}}ek}, \citenamefont {Fiur{\'a}{\v{s}}ek},\
  and\ \citenamefont {Je{\v{z}}ek}}]{Hradil2004}%
  \BibitemOpen
  \bibfield  {author} {\bibinfo {author} {Z.~Hradil}, \bibinfo {author}
  {J.~{\v{R}}eh{\'a}{\v{c}}ek}, \bibinfo {author} {J.~Fiur{\'a}{\v{s}}ek},\
  and\ \bibinfo {author} {M.~Je{\v{z}}ek},\ }in\ \href
  {https://doi.org/10.1007/978-3-540-44481-7_3} {\emph {\bibinfo {booktitle}
  {Quantum state estimation}}}\ (\bibinfo  {publisher} {Springer},\ \bibinfo
  {year} {2004})\ pp.\ \bibinfo {pages} {59--112}\BibitemShut {NoStop}%
\bibitem [{\citenamefont {Maslov}(2017)}]{Maslov2017}%
  \BibitemOpen
  \bibfield  {author} {\bibinfo {author} {D.~Maslov},\ }\emph {Basic circuit
  compilation techniques for an ion-trap quantum machine},\ \href
  {https://doi.org/10.1088/1367-2630/aa5e47} {\bibfield  {journal} {\bibinfo
  {journal} {New Journal of Physics}\ }\textbf {\bibinfo {volume} {19}},\
  \bibinfo {pages} {023035} (\bibinfo {year} {2017})}\BibitemShut {NoStop}%
\bibitem [{\citenamefont {Ryan-Anderson}(2018)}]{ryan2018quantum}%
  \BibitemOpen
  \bibfield  {author} {\bibinfo {author} {C.~Ryan-Anderson},\ }\emph {\bibinfo
  {title} {Quantum Algorithms, Architecture, and Error Correction}},\
  \href@noop {} {Ph.D. thesis},\ \bibinfo  {school} {The University of New
  Mexico} (\bibinfo {year} {2018})\BibitemShut {NoStop}%
\end{thebibliography}%
\end{document}